\documentclass[
reprint,
superscriptaddress,
altaffilletter,
nofootinbib,
amsmath,
amssymb,
aps,
prd,
]{revtex4-2}

\usepackage{enumerate} 
\usepackage{array}
\newcolumntype{P}[1]{>{\centering\arraybackslash}p{#1}}
\usepackage{times}
\usepackage{mathtools}
\usepackage{graphicx}
\usepackage{dcolumn}
\usepackage{bm}
\usepackage{verbatim} 
\usepackage{cprotect} 
\usepackage{url}
\usepackage{tikz}
\usepackage{afterpage}
\usepackage{nicefrac}
\usepackage{multirow}
\usepackage[english]{babel}
\usepackage{natbib} 
\usepackage{glossaries} 
\usepackage{hyperref}
\usepackage{aas_macros} 
\usepackage{comment}
\usepackage{physics}
\usepackage[caption=false]{subfig}

\graphicspath{{figures/}}

\usepackage{xcolor}

\begin{document}

\title{Field localisation and spin-momentum locking in zero-dimensional dissipative topological photonic interface state}

\author{Aidan H.Y. Chong}
\affiliation{Department of Physics, The Chinese University of Hong Kong, Hong Kong}

\author{Y.Q. Liu}
\affiliation{Department of Physics, The Chinese University of Hong Kong, Hong Kong}

\author{C. Liu}
\affiliation{Department of Physics, The Chinese University of Hong Kong, Hong Kong}

\author{Daniel H.C. Ong}
\email{hcong@phy.cuhk.edu.hk}
\affiliation{Department of Physics, The Chinese University of Hong Kong, Hong Kong}

\date{\today}


\begin{abstract}
\noindent Topological photonic systems support edge states that are robust against disorder and perturbation.  Depending on the symmetry and dimensionality of the bulk systems, different edge states emulating soliton, quantum integer and quantum spin Hall effects have been realized.  A major concern in photonics is how one can shape the strength and polarisation of electromagnetic fields to suit different applications.  Here, we show zero-dimensional (0D) interface state arising from one-dimensional (1D) dissipative topological photonic crystals exhibit strong field localisation and spin-momentum locking thanks to its complex classical analogue Dirac mass parameter.  By using spatiotemporal coupled mode theory to formulate 1D photonic crystals and their corresponding Jackiw Rebbi-like (JR) interface state, we find the interaction between two energy bands at high symmetry points plays a major role in defining not only the topological triviality of the crystals but also its complex Dirac mass parameter.  More importantly, when two topological trivial and nontrivial bulk systems are brought together to form a JR state, while the real part of the Dirac mass parameter governs the spectral and spatial field localisations of the interface state, the imaginary part gives rise to a net flow of energy towards the interface and a transverse spin angular momentum, resulting in a strong spin-momentum locking.  We verify our theory by 1D plasmonic crystals using finite-difference time-domain simulations as well as far-field angle-resolved spectroscopy and imaging.
\end{abstract}

\maketitle


\section{Introduction}

\noindent The success of topological insulators in condensed matter physics~\cite{Moore_2010, annurev-conmatphys-062910-140432, RevModPhys.83.1057} has sparked an intensive interest in transferring the concept of topology to electromagnetism~\cite{Lu_Joannopoulos_Soljačić_2014, science.aar4003, science.aar4005, Xie:18, Khanikaev_Alù_2024}, mechanics\cite{RevModPhys.96.021002, Huber_2016}, acoustic~\cite{He_Ni_Ge_Sun_Chen_Lu_Liu_Chen_2016, He_Lai_He_Yu_Xu_Lu_Chen_2020, Chen_Chaunsali_Christensen_Theocharis_Yang_2021}, and electronics~\cite{Gilbert_2021, sciadv.adz2408, Lee_Imhof_Berger_Bayer_Brehm_Molenkamp_Kiessling_Thomale_2018}. Many intriguing properties in topological insulators such as bulk-edge correspondence~\cite{Li_Hu_2023}, spin-momentum locking~\cite{Luo_He_Li_2017}, and quantum spin and anomalous Hall effects~\cite{science.1167733, Hsieh_Qian_Wray_Xia_Hor_Cava_Hasan_2008, PhysRevLett.130.196401} found in topological insulators have been demonstrated to have equivalences in the classical counterparts~\cite{PhysRevB.94.205105, Afzal:18, sciadv.aaw4137, PhysRevB.111.075304}.  Such new horizons are expected to revolutionize the designs and applications of photonic, mechanical, acoustic, and electrical devices. In addition, unlike the topological insulators where energy consumption is mostly considered lossless or Hermitian, classical systems are inherently lossy and leaky, subject to absorption and radiation losses~\cite{Nasari:23, science.aay1064, Midya_Zhao_Feng_2018}. These dissipations manifest many peculiar effects including non-Hermitian skin effect~\cite{PhysRevLett.124.086801}~\cite{Cai_Wang_Zhang_Liu_Nori_2025}, exceptional points~\cite{Yi_Wang_Shi_Wan_Tang_Hu_Li_Zeng_Song_Li_2025a, science.aar7709, Xu:17}, and parity-time symmetry~\cite{Fritzsche_Biesenthal_Maczewsky_Becker_Ehrhardt_Heinrich_Thomale_Joglekar_Szameit_2024, Xu:17, science.abf6873} that are also strongly associated with the topology of the system.  As a result, classical topological systems on the one hand share many similarities with topological insulators while on the other hand are highly distinctive from them.\\

\noindent Recently, topological photonics has gone beyond conventional signal waveguiding and into quantum~\cite{Gao_Xu_Yang_Zwiller_Elshaari_2024, 10.1063/5.0239265} and nonlinear optics~\cite{10.1063/1.5142397, Szameit_Rechtsman_2024} where light-matter interaction is of importance.  In fact, the edge state interacts favourably with quantum dots, molecular dyes, and nonlinear optical materials for spontaneous emission enhancement~\cite{Zeng:19}, lasing~\cite{science.aar4003, science.aar4005}, harmonic generations~\cite{Di_Gaspare_Ghayeb_Zamharir_Knox_Yagmur_Sasaki_Salih_Li_Linfield_Freeman_Vitiello_2025}, and wave mixings~\cite{Dong:24}.  To maximize such interactions, the characteristics of the electromagnetic fields, such as the field strength and the helicity, of the edge states should be optimized properly.  For example, for enhanced emission and harmonic generations, a small mode volume is essential for facilitating strong field strength~\cite{Inada_Hashiya_Nitta_Tomita_Tsujimoto_Suzuki_Yamaki_Hirasawa_2016, PhysRevLett.110.237401, 10.1063/1.5064468, PhysRevResearch.4.023233}.  In addition to the field localisation, the polarisation of the electric fields is also of interest.  In analogy to the electron spin, the helicity of the fields mediates the chiral light-matter interface and spin-momentum locking, which play a dominant role in spintronics~\cite{Kohda_Okayasu_Nitta_2019}, topological quantum computing~\cite{science.1231473}, and optically controlled devices~\cite{10.1063/1.3111442}.\\

\noindent Most of the studies on topological photonics have been focusing on realizing edge state with different dimensionalities under various bulk symmetries.  For one-dimensional (1D) photonic systems that satisfy inversion symmetry, much effort is reported to emulate the Su-Schrieffer-Heeger (SSH) model for constructing a zero-dimensional (0D) Jackiw-Rebbi-like (JR) state at the interface state between two 0 and $\pi$ Zak phases systems~\cite{Dirac-eqt-tutorial, PhysRevB.106.045401, PhysRevResearch.4.023233}.  For two-dimensional (2D) honeycomb systems, using the so-called expanded and shrunken deformed lattices to construct pseudo time-reversal symmetry as well as pseudospin states for rendering Kramers’ degeneracy, the classical analogue of quantum spin Hall effect has been demonstrated in which spin-momentum locking clearly observed from the 1D edge state~\cite{Yang_Jiang_Hang_2018, Shao:23}.  Other than honeycomb lattices, 2D square lattices that support both time-reversal and inversion symmetries can render 2D vectorial Zak phase facilitating as topological invariant, forming not only 1D edge state but also higher order 0D corner state~\cite{PhysRevLett.122.233902, acsphotonics.2c00571}.  However, for all these edge states, not much work is devoted to studying their field properties.  For example, 1D photonic crystals, although it is well known that varying the grating fill factor can lead to 0 and $\pi$ Zak phases, how it affects the field properties of the consequent JR state remains uncertain.  In fact, the knowledge of tailoring the energy, propagation, strength, and polarisation, of the edge sate is crucial for engineering the light-matter interaction.\\

\noindent In this work, we formulate the electric fields as well as the transverse spin angular momentum (SAM) of the Jackiw Rebbi (JR) state in photonic systems by using spatiotemporal coupled mode theory (CMT)~\cite{Bykov:15}.  We apply CMT at high symmetry points of the bulk systems where two energy bands couple and then transform it to the SSH model and the Dirac equation in 1+1 dimensions.  Through the transformations, we find the complex coupling constant between the bands acts equivalently to Dirac mass, which defines the topology of the system.  More importantly, when two topological trivial and nontrivial bulks are brought together, their respective complex Dirac masses define the overall electric field profiles of the JR state.  While the real part of the Dirac mass governs the spectral and spatial localisation of the evanescent fields at the interface, the imaginary part gives rise to a propagation term that drives energy flowing towards the interface.  Such propagation inherently connects with a transverse SAM, thus manifesting strong spin-momentum locking around the interface.  We verify our theory by 1D plasmonic crystals (PmCs) using finite-difference time-domain simulations and far-field angle-resolved spectroscopy and imaging.\\


\section{Theory}\label{section: Theory}

\subsection{Two-Band Model Topology by Inversion Symmetry and Dirac Approximation}

\begin{figure}[h!]
\centering
\includegraphics[width=1\linewidth]{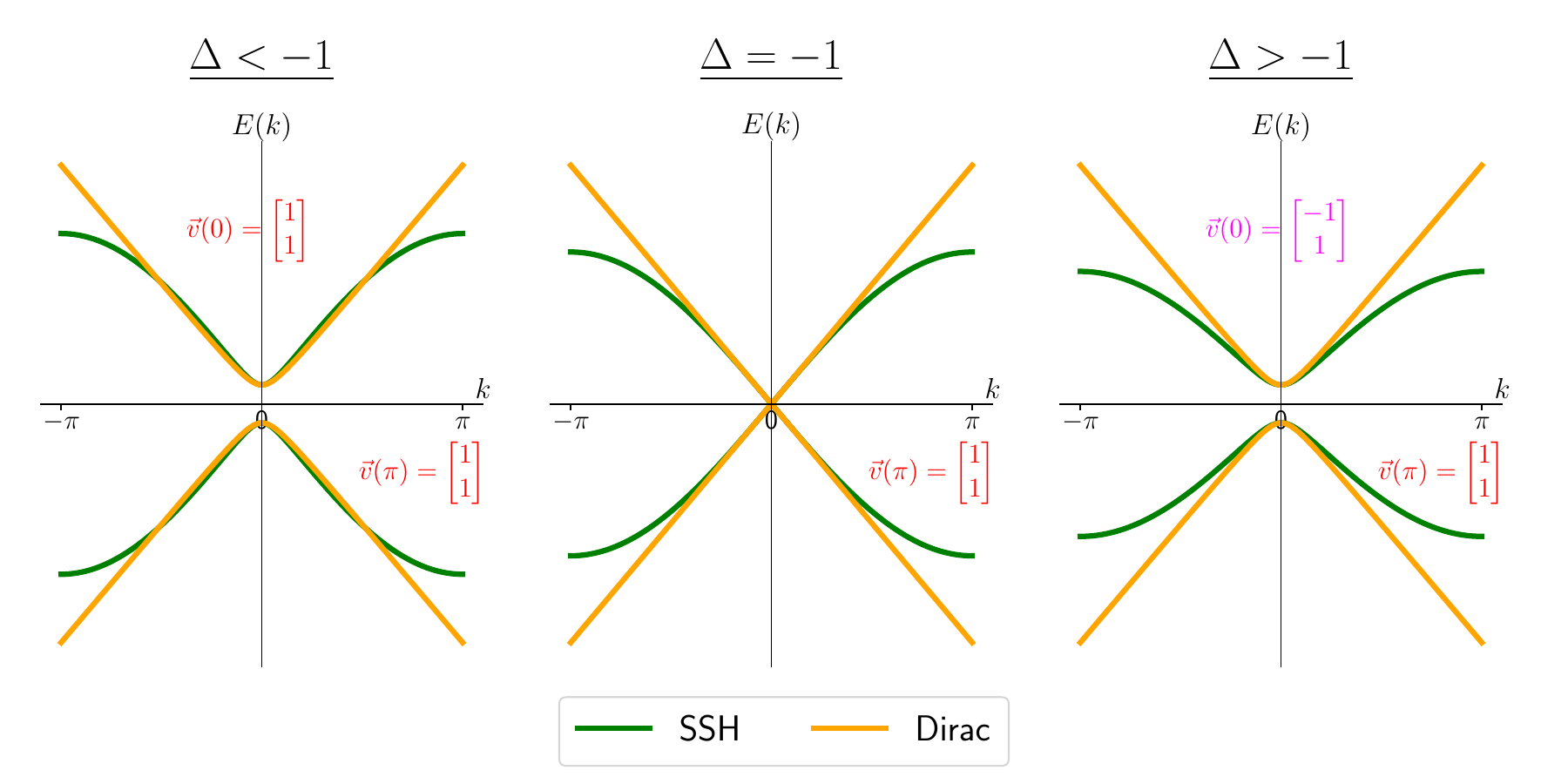}
\caption{Figure showing the eigenenergy $E(k)$ of the SSH model plotted as a function of momentum $k$ in the first Brillouin zone for 3 different cases. As the value of $\Delta$ changes from $<-1$ to $>-1$, eigenvector $\vec{v}(\pi)$ remains unchanged. The inversion eigenvalue at $\pi$ stays at $\lambda(\pi)=1$. However, the eigenvector $\vec{v}(0)$ changes from $\vec{v}(0)=^T$ to $\vec{v}(0)=[-1,1]^T$ after $\Delta$ passes the Dirac point $\Delta=-1$. }
\label{fig: SSH and Dirac}
\end{figure}

\noindent All one-dimensional, two-band insulators without symmetries are topologically trivial~\cite{Dirac-eqt-tutorial}. Inversion symmetry is one of the simplest crystalline symmetries that can produce topological phases. We focus on a system whose topology is protected solely by inversion symmetry. For a two-band model, the Bloch Hamiltonian commutes with the inversion operator at the high-symmetry points (HSPs) $k=0$ and $k=\pi$, i.e., $[\mathcal{H}(k), \mathcal{I}]=0$. Here $\mathcal{H}(k)$ is the Bloch Hamiltonian and $\mathcal{I}=\sigma_x$ is the inversion operator. Under this condition, the eigenstates of $\mathcal{H}$ at the HSPs can also be chosen as eigenstates of $\mathcal{I}$. The inversion eigenvalues are quantized to $\lambda(k)=\pm 1$ at the HSPs. Examining the lower band, the state is topological when $\lambda(0) = -\lambda(\pi)$ and trivial when $\lambda(0) = \lambda(\pi)$. The inversion-symmetric Bloch Hamiltonian of the Su-Schrieffer-Heeger (SSH) model is~\cite{Dirac-eqt-tutorial}\cite{A_short_course_on_topological_insulator}
\begin{equation}
\mathcal{H}_{\text{SSH}}(k) = (v + w\cos(k))\sigma_x + w\sin k \sigma_y.
\label{SSH model}
\end{equation}
While $v$ and $w$ are generally complex and carry phases, these phases can always be removed by a gauge transformation\cite{A_short_course_on_topological_insulator}. Typically, $v$ and $w$ are taken to be real and positive, in which case the band gap closes at the zone boundary $k=\pi$. For our case, we select a gauge such that a single band gap closes at $k=0$ in the first Brillouin zone. This condition is achieved by setting $w=1$ and $\Delta = v/w$, where $v$ is a negative real number. Considering $\mathcal{H}_{\text{SSH}}(k)$ for $\Delta \in (-\infty,0)$, the topological phase exists for $\Delta<1$, while the system is trivial otherwise. A single band gap closes at $k=0$ when $\Delta=-1$ (see Fig. \ref{fig: SSH and Dirac}).\\

\noindent By evaluating the eigenvectors at both HSPs ($k=0,\pi$), we find that the inversion eigenvalues differ only at $k=0$: $\lambda(0) = -1$ in the topological phase and $\lambda(0) = 1$ in the trivial phase, while $\lambda(\pi)=1$ remains constant across the phase transition (see Fig.\ref{fig: SSH and Dirac}). Therefore, the eigenvectors and inversion eigenvalues change only at $k=0$ and not at $k=\pi$. Consequently, it is sufficient to understand the topological properties of the entire system by examining the behavior near $k=0$. A Taylor expansion of Eq.\ref{SSH model} at $k=0$ yields the 1+1D Dirac equation ($c=\hbar=1$):
\begin{equation}
\mathcal{H}_{\text{Dirac}} = m \sigma_x + k \sigma_y,
\label{Dirac equation}
\end{equation}
where we define $m = \Delta+1$. Thus $m>0$ corresponds to the topological phase, while $m<0$ corresponds to the trivial phase.

\subsection{Spaciotemporal Coupled-Mode Theory (CMT)}

\noindent We first consider a one-dimensional (1D), optically thick plasmonic crystal (PmC) that supports Bloch-like surface plasmon polaritons (SPPs) along the $\Gamma$-X direction. The coordinate system is defined in Fig.\ref{fig: schematic diagram and SEM}(a), and it is used for the rest of the paper. The grating vector is $k_0 = \frac{2 \pi n}{P}$, where $n$ is the Bragg scattering order and $P$ is the period of the PmC. Near normal incidence, two counter-propagating SPPs follow the linearized dispersion relation $\tilde{\omega} \approx \tilde{\omega_0} \pm v_gk_x$. Here $k_x$ is the propagation constant, $\tilde{\omega}$ is the complex angular frequency of the mode, $\tilde{\omega_0}$ is the complex angular frequency at the band intersection (the $\Gamma$ point), and $v_g$ is the group velocity.\\

\noindent The intrinsic optical properties of guided-mode resonant gratings can be described by spatiotemporal coupled-mode theory (CMT)~\cite{Bykov:15, Haus-1984, Wong:18}. The governing CMT equation for two counter-propagating surface plasmons on a 1D grating is
\begin{equation}
\begin{bmatrix} v_g [k_x-i(\partial_xA_1)\frac{1}{A_1}] & \tilde{\omega_c}\\ \tilde{\omega_c} & -v_g [k_x-i(\partial_xA_2)\frac{1}{A_2}]\end{bmatrix} \begin{bmatrix} a_1\\ a_2 \end{bmatrix} = \tilde{\Omega} \begin{bmatrix} a_1\\ a_2 \end{bmatrix}.
\label{general_CMT_eigenequation}
\end{equation}
$A_{1,2}$ represent the amplitudes of the wave packets. $v_g$ is the group velocity. $\tilde{\Omega} = \tilde{\omega}-(\omega_0-i\alpha)$, where $\omega_0$ and $\alpha$ are the uncoupled center frequency and decay constant, respectively. $\tilde{\omega_c} = \omega_c'-i\omega_c''$ is the complex coupling constant. Because the imaginary part is associated with decay, $\omega_c''$ must be positive for the solution to be physical. $k_x$ is the $x$ component of the incident light. $A_{1,2}$ also decay exponentially in the $\hat{z}$ direction, which is implied in the following calculation.\\

\noindent For uniform illumination on a 1D infinite grating (bulk), $A_{1,2}$ are constant. Therefore, the solution to Eqt.\ref{general_CMT_eigenequation} is straightforward. Since the topological phases of such systems can be determined entirely at $k=0$, we find the analytical solutions at normal incidence. The eigenvalues and eigenvectors are 
\begin{equation}
    \tilde{\Omega}_{\pm} = \pm \tilde{\omega_c}, \quad \vec{v_{\pm}}= \begin{bmatrix} \pm 1 \\ 1\end{bmatrix}.
\label{eigenvalues and eigenvectors}
\end{equation}
By analogy with the Dirac equation, where the sign of the mass term determines the topology, we identify the real part of the complex coupling constant $\omega_c'$ as defining the topology. Evaluating Eqt.\ref{eigenvalues and eigenvectors} for $\omega'_c > 0$ and $\omega'_c < 0$ yields eigenvectors identical to those of the Dirac equation (see Fig.\ref{fig: SSH and Dirac}). Consequently, the 1D SPP system is in the topological phase when $\omega_c'>0$ and is trivial otherwise.\\

\noindent The correspondence between the 1D SPP system and the Dirac equation is not coincidental. The SPP Hamiltonian derived from CMT can be unitarily transformed into the Dirac equation. The mass term is mapped to the complex coupling constant by $m = \frac{\tilde{\omega_c}}{v_g}$. Consequently, the two systems are expected to exhibit identical topological properties.\\

\subsection{Non-Hermitian Jackiw-Rebbi State of SPP}

\noindent We seek a bounded solution at the interface between two topologically distinct phases at normal incidence ($k_x=0$), analogous to the JR state of the Dirac equation. In contrast to the Hermitian Dirac equation, the CMT Hamiltonian is non-Hermitian. Specifically, the coupling constant $\tilde{\omega_c}$ is generally complex, whereas the mass term $m$ in the Dirac equation is real. We therefore model an interface between two phases by defining the spatial distribution of the coupling constant $\tilde{\omega_c}(x)$ as a step function. For region $x<0$, we define $\tilde{\omega_c} = \tilde{\omega_1} = -\omega_1'-i\omega_1''$. For region $x>0$, we define $\tilde{\omega_c} = \tilde{\omega_2} = \omega_2'-i\omega_2''$. Here all variables $\omega_{1,2}'$ and $\omega_{1,2}''$ are real positive constants. Physically, this model represents two semi-infinite gratings, potentially with different coupling constants $\tilde{\omega_c}$, joined at $x=0$ to form an interface. For simplicity, we assume the uncoupled frequency $\tilde{\omega_0} = \omega_0' - i\alpha$ is the same on both sides of the grating. These parameters are illustrated in Fig.~\ref{fig: schematic diagram and SEM}(a). We seek a bounded solution localized at the interface, which requires the wavefunction to decay to zero as $x \rightarrow \pm\infty$ (Dirichlet boundary condition).\\

\noindent By determining the eigenvalues and eigenfrequencies on both sides and matching the boundary condition at $x=0$, a solution is found at $\tilde{\Omega} = 0$. If $\tilde{\omega_1}=\tilde{\omega_2}$, the solution reduces to the bulk solution, which verifies the analytical result. The normalized non-Hermitian Jackiw-Rebbi solution is 
\begin{equation} 
\begin{split}
\begin{bmatrix} \psi^+(x) \\\psi^-(x) \end{bmatrix} = \mathcal{N} \begin{bmatrix} i \\1 \end{bmatrix} e^{-\frac{\tilde{\omega_c}}{v_g}x} &= \mathcal{N}\begin{bmatrix} i \\1 \end{bmatrix} e^{-\tilde{m}x},\\
\mathcal{N} = \sqrt{\frac{2}{v_g}\left(\frac{\omega_1' \omega_2'}{\omega_1' + \omega_2'}\right)} &= \sqrt{2\left(\frac{m_1' m_2'}{m_1' + m_2'}\right)},
\end{split}
\label{Non-Hermitian Jackiw-Rebbi solution} 
\end{equation} 
where the superscripts $+$ and $-$ denote the traveling direction of the SPP wave along the $\hat{x}$ direction, and $\mathcal{N}$ is the normalisation constant. Since the mass term is mapped to a complex number, $\tilde{m}$ is complex and defined as $\tilde{m} = m' + im'' = \frac{\tilde{\omega_c}}{v_g}$. As the sign of $\omega_c'$ changes at $x=0$, the two wave functions decay approximately exponentially away from the interface when $\omega_c''$ is small, with $|\frac{\omega_c'}{v_g}|$ setting the decay constant. 

\subsection{Poynting Vector and Spin Angular Momentum}\label{Poynting and SAM}

\noindent Writing out the magnetic field explicitly, we obtain the analytical form of the magnetic field of the interface state produced by the two counter-propagating SPP waves
\begin{equation}
    \vec{H} = H_o(z,t)e^{-\tilde{m}x}(i e^{i \tilde{k} x}+ e^{-i \tilde{k} x})\hat{y},
\label{H field}
\end{equation}
where $\tilde{k} = k + i\kappa$. The corresponding electric field can be found analytically using Maxwell's equations and is likewise dominated by the exponential decay term $e^{-\tilde{m}x}$.\\

\noindent Our system is dissipative. Radiation is primarily emitted through the interface state, while SPPs are continuously absorbed by the metal during propagation. The time-averaged Poynting vector along the x-direction is $\langle S_x \rangle = \frac{1}{2}\text{Re}[-H_y^*E_z]\hat{x}$, where $H_y^*$ is the complex conjugate of the magnetic field. The net Poynting vector along $\hat{x}$ can be found analytically:
\begin{equation}
\begin{split}
\langle P_x \rangle = P_{0x}(z,t)e^{-2m'x}(&m''[\cosh(2 \kappa x)-\sin(2kx)]\\
-&k\sinh(2\kappa x)-\kappa \cos(2kx))
\end{split}
\label{Px} 
\end{equation} 
Although the spatial decay is largely controlled by the $e^{-2m'x}$ term, the decay terms $m''$ and $\kappa$ play significant roles. As $\omega_c'$ changes sign at $x=0$ while $\omega_c''$ remains positive, the Poynting vector points toward $x=0$ on both sides of the interface. This creates a funneling effect similar to that in leaky photonic crystals~\cite{Yoon2022}. For photonic systems, the only dissipation channel is radiative decay, so the existence of a nonzero Poynting vector depends solely on the imaginary part of the Dirac mass $m''$~\cite{Yoon2022}. If the system is purely Hermitian, then both $m''$ and $\kappa$ are zero and hence $\langle P_x \rangle=0$. Therefore, the funneling effect of the topological interface state can only manifest in a non-Hermitian system.\\ 

\noindent Unlike purely photonic systems, plasmonic systems carry intrinsic spin~\cite{Bliokh2015}. Due to additional decay from metal absorption, the two counter-propagating SPPs can decay at different rates, producing a transverse spin angular momentum (SAM) on each side of the interface state. Using the definition of SAM for light~\cite{vernon2024decomposition}, the transverse SAM of the plasmonic JR state is 
\begin{equation} 
\begin{split}
S_{\perp} = -S_{0}(z,t)e^{-2m'x}(&m''[\cosh(2 \kappa x)-\sin(2kx)]\\
-&k\sinh(2\kappa x)-\kappa \cos(2kx))\hat{y}, 
\end{split}
\label{interface state transverse SAM} 
\end{equation} 
Evidently, both the Poynting vector and transverse SAM share the same functional form. As they decay and change direction in the same way, we conclude that the non-Hermitian plasmonic JR state exhibits spin-momentum locking. This spin-momentum locking is a key result of our work.\\ 

\subsection{Mode Volume} \label{mode volume} 

\noindent The mode volume $V_m$ is a key figure of merit used to quantify the spatial confinement of electromagnetic energy. While several definitions exist, we adopt the expression given in ~\cite{Srinivasan:06}. Since minimizing the mode volume corresponds to maximizing the electromagnetic energy concentration, we seek the conditions that minimize $V_m$. From Eqt.\ref{Non-Hermitian Jackiw-Rebbi solution}, the field decays approximately as $\propto e^{-m'x}$. The inverse mode volume $V_m^{-1}$ can be found analytically using the normalisation constant $\mathcal{N}$: 
\begin{equation} \frac{1}{V_m} = \frac{2}{v_g}\frac{\omega_1'\omega_2'}{\omega_1' +\omega_2'}, 
\label{dual sided V} 
\end{equation} 
which depends only on the real parts of the coupling constants, $\omega'_{1,2}$. Consequently, the mode volume $V_m$ is minimized by maximizing the coupling constants $\omega'_{1}$ and $\omega'_{2}$, which are related to the photonic band gaps of the respective media.\\ 

\noindent In summary, we have derived the topological properties of a 1D plasmonic system using the spatiotemporal CMT framework. We identified the mathematical equivalence between the CMT Hamiltonian and the massive 1+1D Dirac Hamiltonian, demonstrating that they exhibit the same topological properties. A zero-energy, non-Hermitian interface state was constructed using CMT with a complex coupling constant $\tilde{\omega_c}$. This interface state is characterized by an exponentially decaying envelope, similar to its Hermitian counterpart. Due to the non-Hermitian nature, our plasmonic JR state exhibits spin-momentum locking. Finally, by analyzing the mode volume of these states, we determined that it can be minimized by maximizing the band gaps of the constituent plasmonic crystals, providing a design principle for concentrating electromagnetic energy in a plasmonic interface state.\\


\section{Finite-Difference Time-Domain (FDTD) Simulation}\label{Section: FDTD}

\noindent We performed Finite-Difference Time-Domain (FDTD) simulations to numerically validate the key predictions of our theoretical model. First, we simulated bulk gratings to demonstrate the bulk inversion symmetry and topological properties. Then we simulated interface states to verify the spatial field decay, Poynting vector, and transverse SAM. Lastly, we used both bulk and interface-state simulations to confirm the optimisation method for mode volume. 

\begin{figure}[h!]
    \centering
    \subfloat[\centering Schematic Diagram]{{\includegraphics[width=0.4\textwidth]{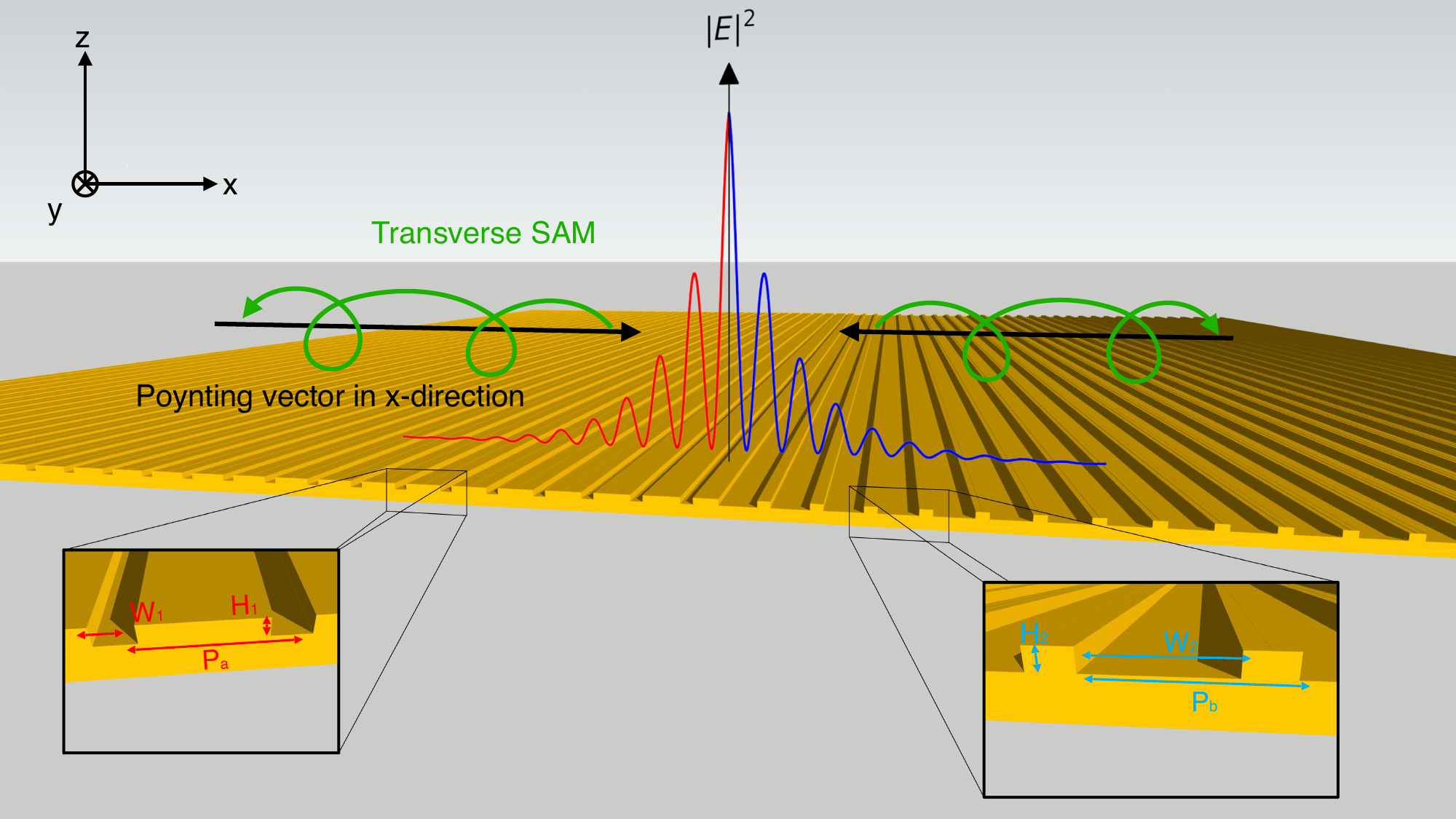}}}
    \qquad
    \subfloat[\centering SEM Image]{\includegraphics[width=0.4\textwidth]{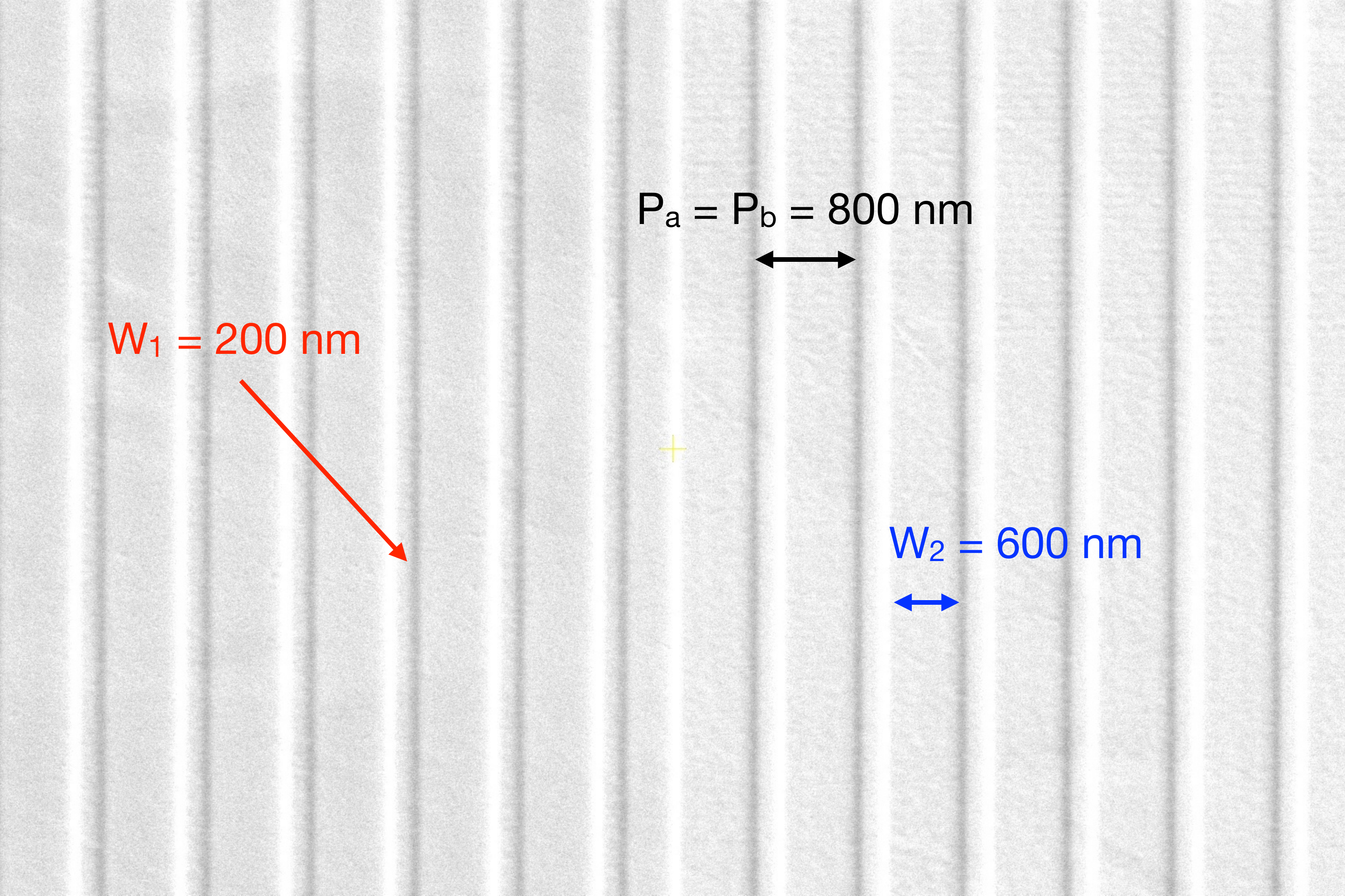}}
    \caption{(a): Schematic diagram showing the different parameters of an interface state. $W_{1,2}$ are the slit widths, $H_{1,2}$ are the slit heights, and $P_{a,b}$ are the grating periods. We choose gold as the material of the gratings for surface plasmon propagation with moderate absorption. In all simulations and experiments, we fixed $H_{1,2}=30$nm and $P_{a,b}=800$nm. By varying $W_{1,2}$ we effectively change $\tilde{\omega}_{1,2}$ in our simulations and experiments. The diagram also shows the spatial decay of the electric field predicted by our theory and verified using simulations and experiments. As our interface state exhibits spin-momentum locking, the x-component of the Poynting vectors points toward the center of the interface state, while the directions of the SAM arise from the propagating surface plasmon. (b) SEM image of FIB-fabricated PmCs forming an interface state. Both gratings have a period of 800nm. The grating on the left has a slit width of 200nm, while the grating on the right has a slit width of 600nm.}
\label{fig: schematic diagram and SEM}
\end{figure}

\subsection{Band Inversion Symmetry}\label{subsection: sim band inversion}

\noindent First, we simulated the band structures of bulk PmCs to confirm the predicted band inversion symmetry. The simulations modeled a series of 1D rectangular Au gratings with fixed period P = 800nm, height H = 30nm, and slit widths W ranging from 100nm to 700nm in 50nm increments. A schematic of the grating geometry is shown in Fig.\ref{fig: schematic diagram and SEM}(a), and the full list of simulated widths is provided in Table \ref{table: width_table}.\\

\begin{table}[h!]
\centering
\begin{tabular}{|c||c|}
    \hline
    \multicolumn{2}{|c|}{Widths W (nm)} \\
    \hline
    100 nm & 700 nm \\
    \hline
    150 nm & 650 nm \\
    \hline
    200 nm & 600 nm \\
    \hline
    250 nm & 550 nm \\
    \hline
    300 nm & 500 nm \\
    \hline
    350 nm & 450 nm \\
    \hline
\end{tabular}
\caption{Each entry represents a bulk grating with the specified width $W$ (12 in total). Each row corresponds to an interface state simulated in Section \ref{subsection: sim interface field}, formed by joining the two listed bulk gratings.}
\label{table: width_table}
\end{table}

\begin{figure}[h!]
\centering
    \centering
    \subfloat[\centering W=200nm]{{\includegraphics[width=0.22\textwidth]{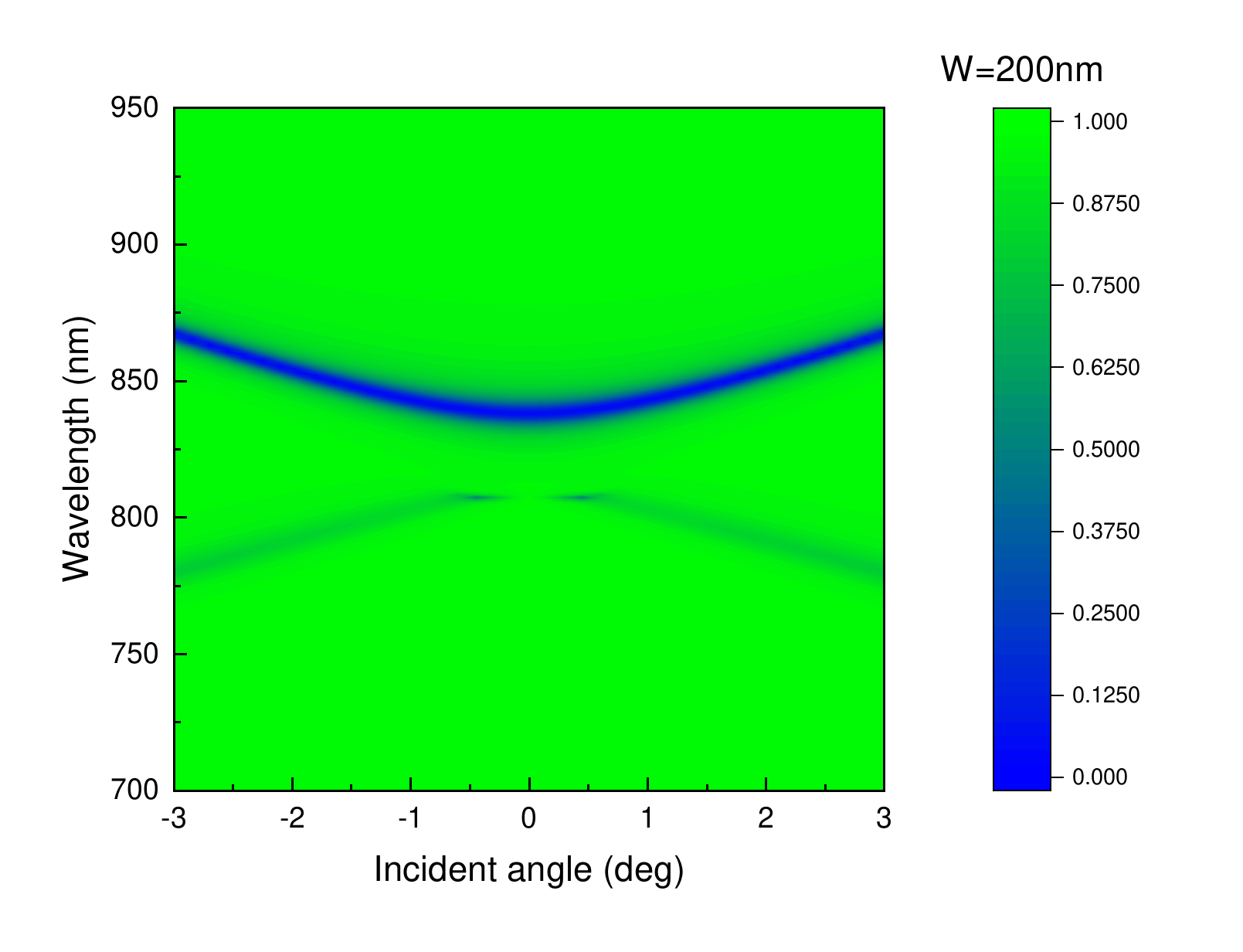}}}
    \subfloat[\centering W=400nm]{{\includegraphics[width=0.22\textwidth]{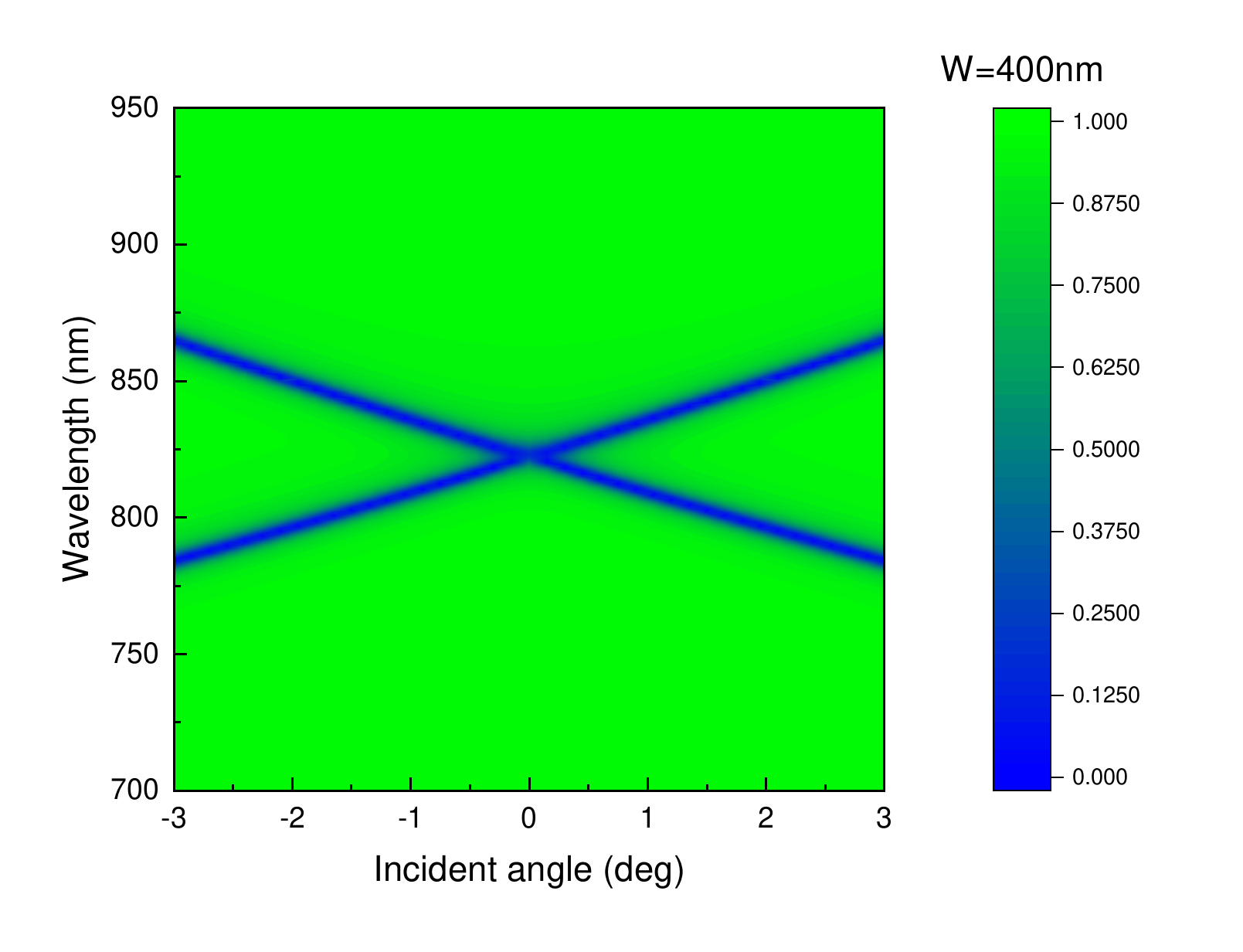}}}
    \qquad
    \subfloat[\centering W=600nm]{{\includegraphics[width=0.22\textwidth]{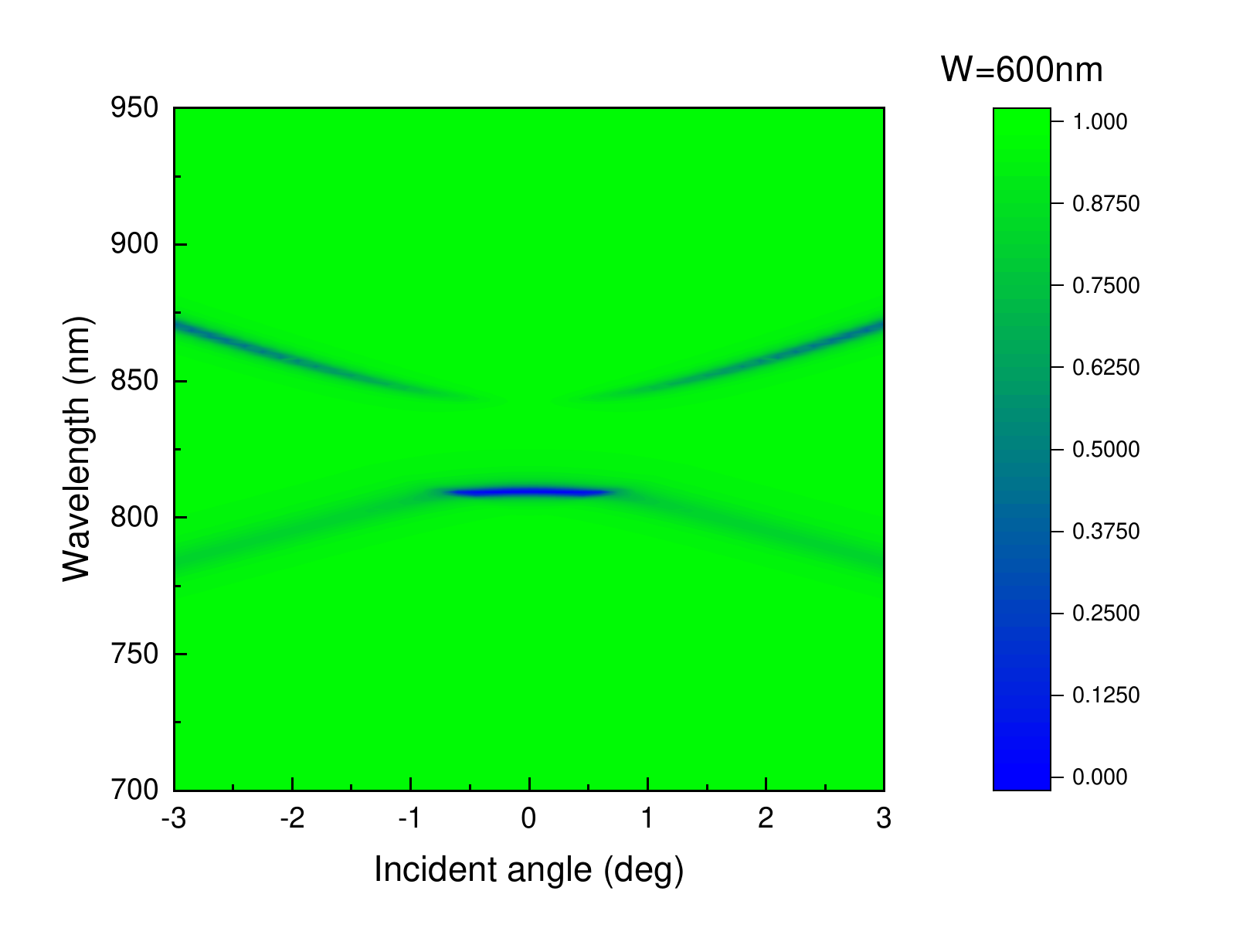}}}
    \subfloat[\centering W=200\&600nm]{{\includegraphics[width=0.22\textwidth]{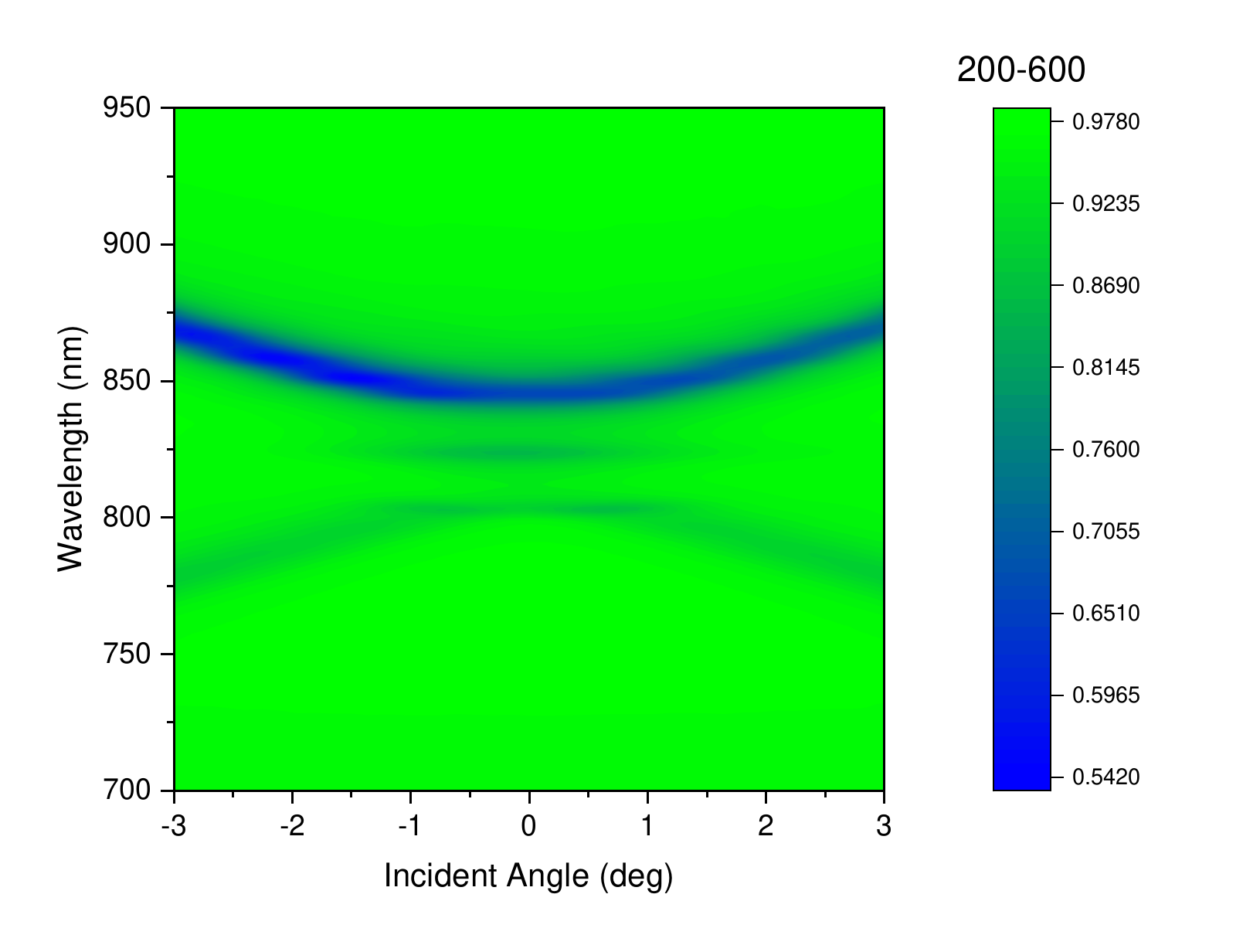}}}
    \qquad
    \subfloat[\centering W=200nm]{{\includegraphics[width=0.22\textwidth]{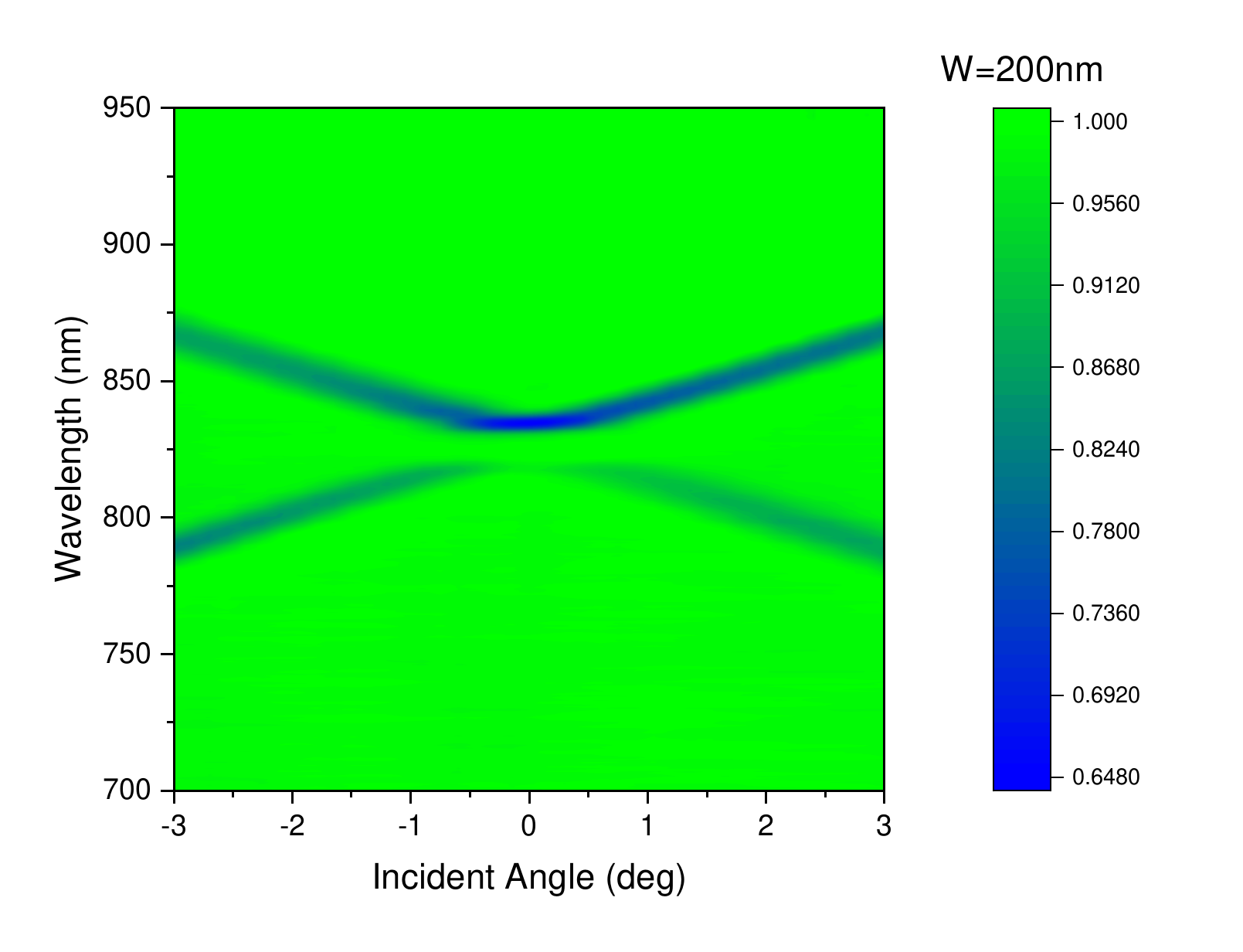}}}
    \subfloat[\centering W=400nm]{{\includegraphics[width=0.22\textwidth]{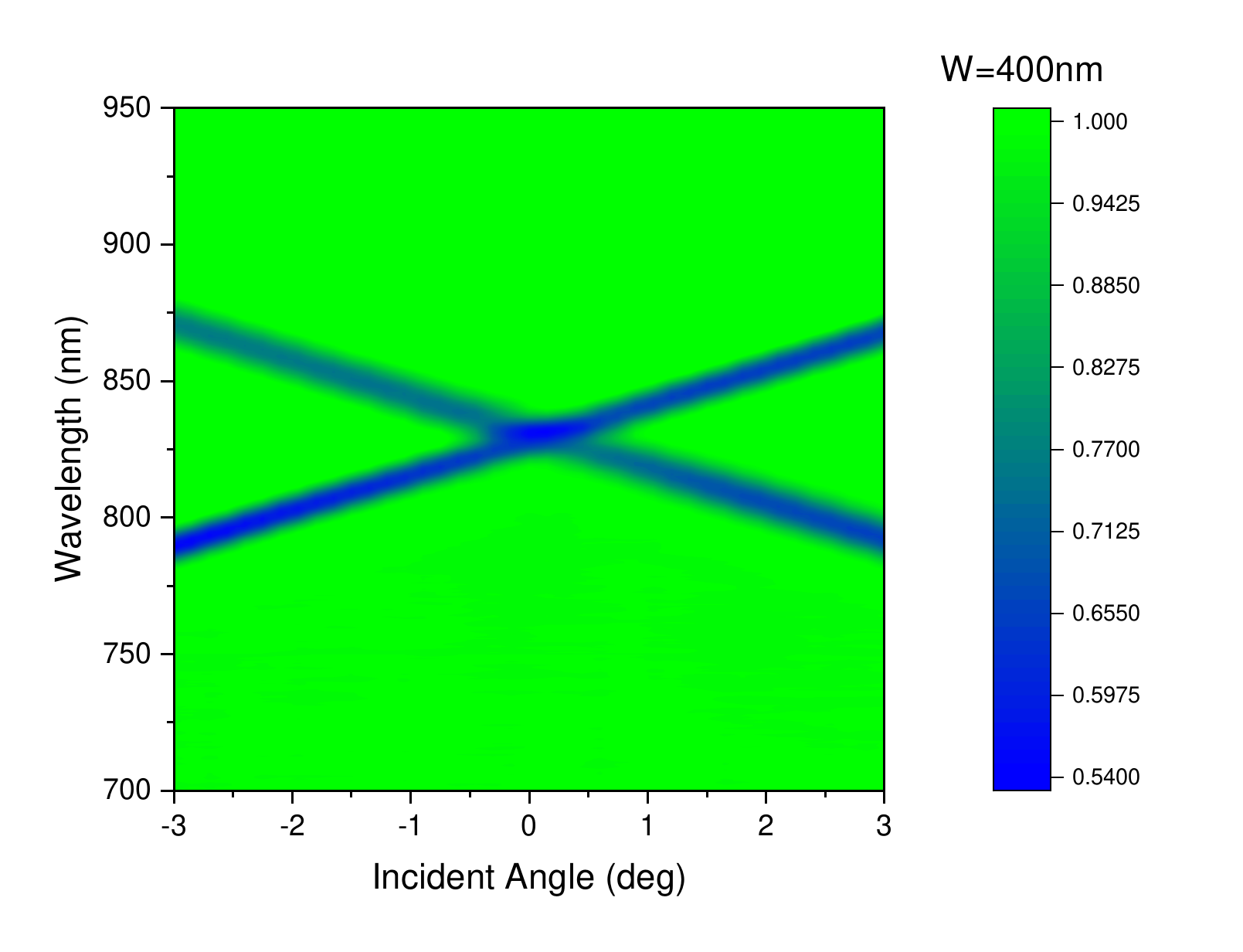}}}
    \qquad
    \subfloat[\centering W=600nm]{{\includegraphics[width=0.22\textwidth]{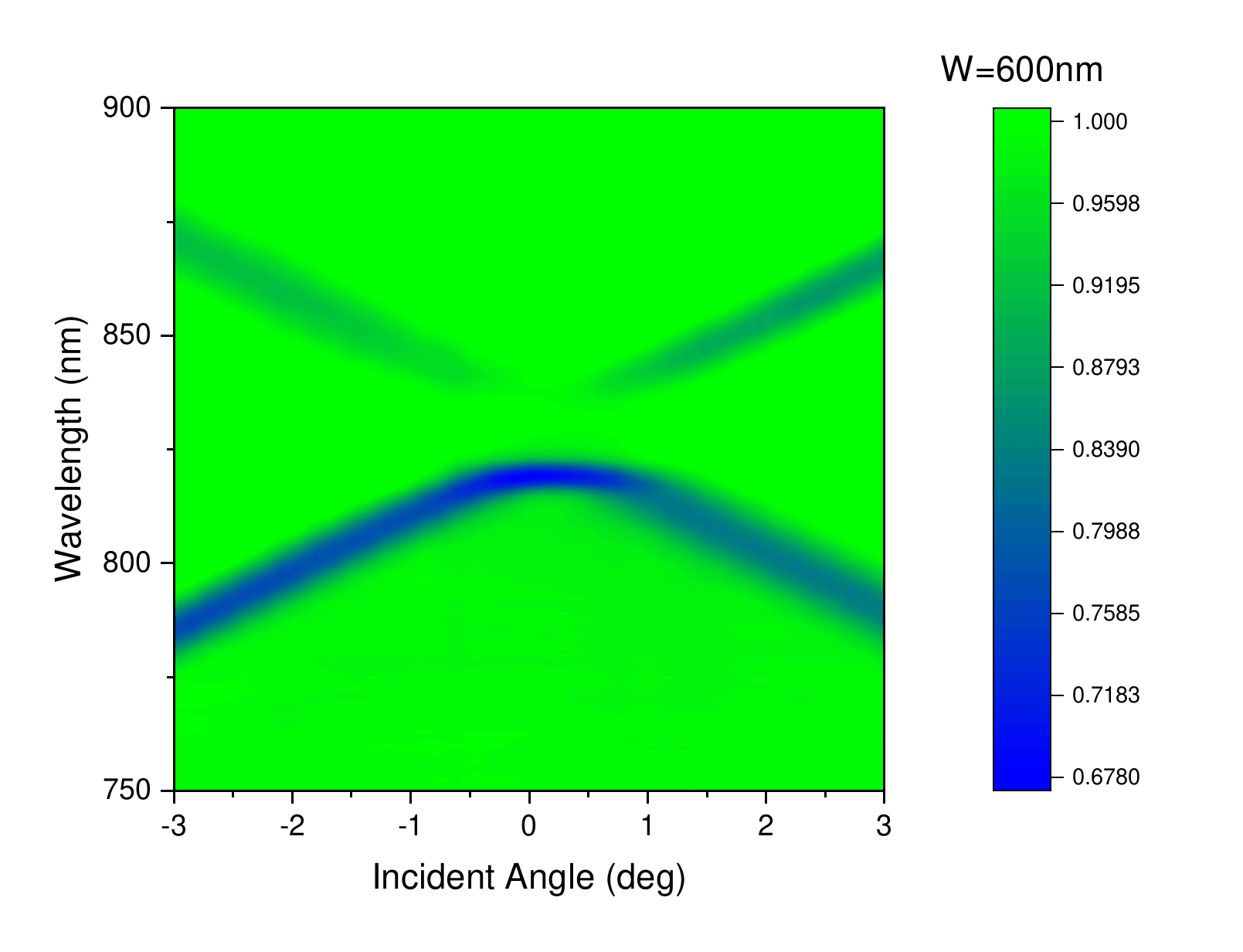}}}
    \subfloat[\centering W=200\&600nm]{{\includegraphics[width=0.22\textwidth]{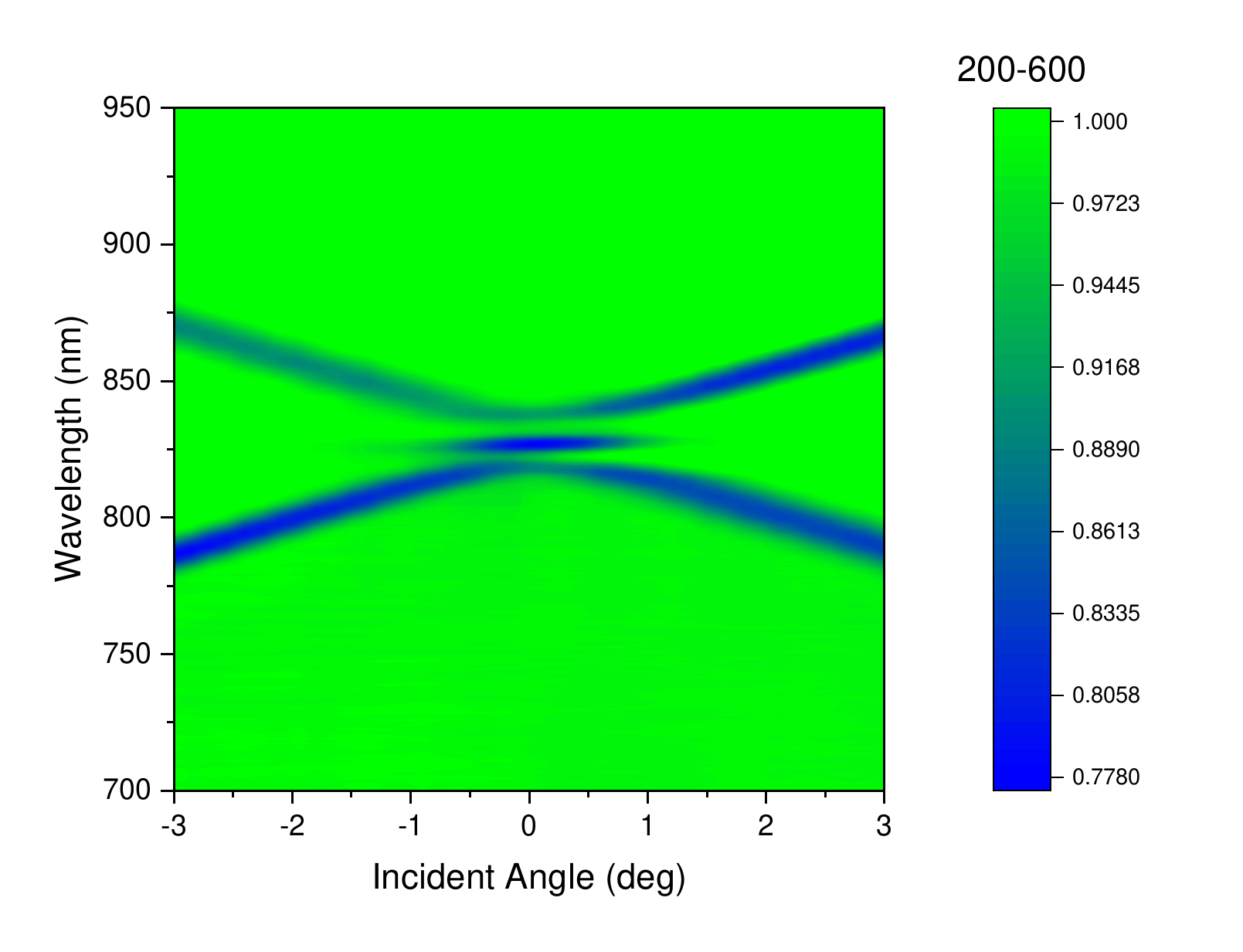}}}
\caption{(a)-(d): Simulated band structures. P-polarised light is incident on the grating at angles ranging from $-3^{\circ}$ to $3^{\circ}$ along the $\Gamma$-X direction. The reflectivity profiles were collected to form a band structure. The $(0,\pm1)$ SPP modes exhibit band inversion symmetry for $W<400$nm and $W>400$nm. The band gap obtained at normal incidence is used to calculate the coupling constant $\omega_c'$ via Eqt.\ref{eigenvalues and eigenvectors}. (e)-(h): Measured band structures of the fabricated PmCs. The experimental band structures are similar to the simulation results in (a)-(d). As with the simulations, $\omega_c'$ is determined from the measured band gaps. (d) and (h) show the band structure when an interface state is formed using $W=200$nm and $W=600$nm gratings. The JR state appears at the center of the band gap.}
\label{fig: band_structures}
\end{figure}

\noindent For the bulk band-structure simulations, a single unit cell was modeled in 2D, which is sufficient because only p-polarized light excites the SPP modes. A plane wave was incident from the top ($+y$ direction), with perfectly matched layers (PMLs) terminating the simulation domain at the bottom ($-y$ direction). Bloch boundary conditions were applied along the x-axis to represent an infinite periodic structure. The incident angle was swept from $-3^{\circ}$ to $3^{\circ}$ in $0.15^{\circ}$ increments, and the resulting reflectivity spectra were compiled to construct the band structures. Representative band structures are shown in Fig.\ref{fig: band_structures}(a-d), which clearly demonstrate the predicted inversion symmetry of the SPP modes. The band gap, $2\omega_c'$, was extracted from the reflectivity spectrum at normal incidence for each grating. The resulting values of $\omega_c'$ as a function of slit width are plotted in Fig.\ref{fig: omega_c_and_dirac_mass}(a).

\subsection{Interface-State Field Pattern Simulation}
\label{subsection: sim interface field}

\noindent Having verified the band inversion symmetry of our system in Section \ref{subsection: sim band inversion}, we next investigated the spatial field decay of the interface states. Specifically, we aimed to validate the relationship between mode confinement and system parameters to establish a design principle for minimizing the mode volume $V_m$.\\ 

\noindent We simulated the interface states formed by the pairs of grating widths listed in Table \ref{table: width_table}. These pairs were chosen to have different topological phases but nearly identical band gaps ($\omega_1' \approx \omega_2'$). For each pair, the near-field electric field intensity $|E(x)|^2$ was recorded at the interface resonance frequency $\omega_0$. Assuming the decay is dominated by the exponential term, the natural logarithm of the field intensity yields a linear relation:
\begin{equation}
    \ln(|E(x)|^2) \propto (-2\frac{\tilde{\omega}_c}{v_g})x\propto (-2m')x.
\label{log_E^2}
\end{equation}
Thus $\omega_c'$ can be extracted directly from a linear fit to the slope of the simulated $\ln(|E(x)|^2)$ data. The simulated near-field patterns and the corresponding linear fits are shown in Fig.\ref{fig: interface state fields}(a).\\

\noindent Our model provides two independent methods for determining the coupling constant $\omega_c'$: from the bulk band gap and from the spatial decay of the interface-state field (Eqt.\ref{log_E^2}). Fig.\ref{fig: omega_c_and_dirac_mass}(a) compares the values of $\omega_c'$ obtained via these two methods for all pairs of simulated gratings. The excellent agreement between the datasets validates the self-consistency of our CMT framework. Using these two sets of $\omega_c'$ values, we calculated the theoretical inverse mode volume $1/V_m$. The results, plotted in Fig.\ref{fig: mode-volume}(a), show that both methods yield nearly identical values for $1/V_m$, confirming that a larger band gap leads to stronger energy confinement.\\

\subsection{Poynting Vector and Spin Angular Momentum} \label{subsection: Poynting and SAM}

\begin{figure}[h!]
\centering
    \centering
    \subfloat[\centering Poynting vector]{{\includegraphics[width=0.25\textwidth]{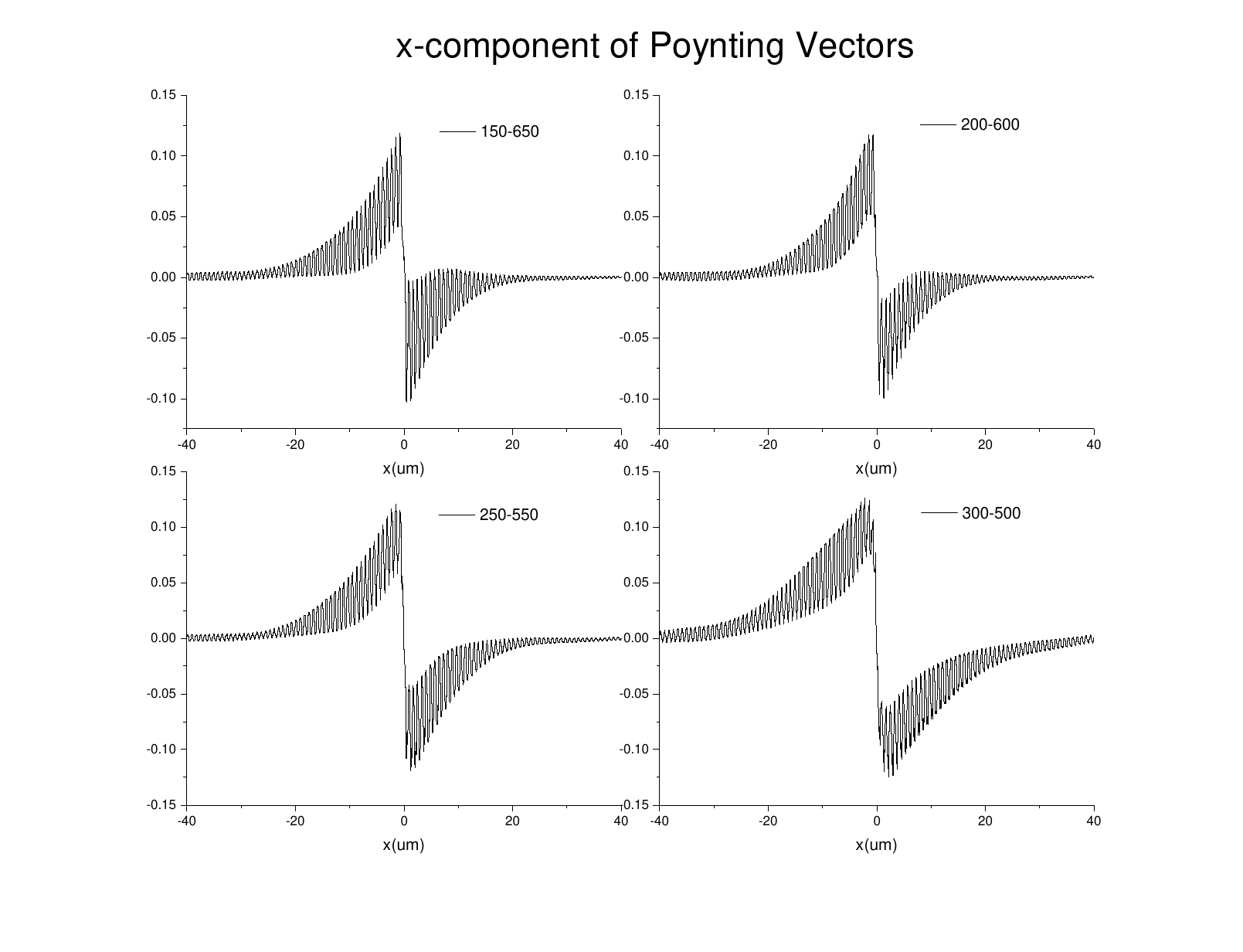}}}
    \subfloat[\centering Transverse SAM]{{\includegraphics[width=0.25\textwidth]{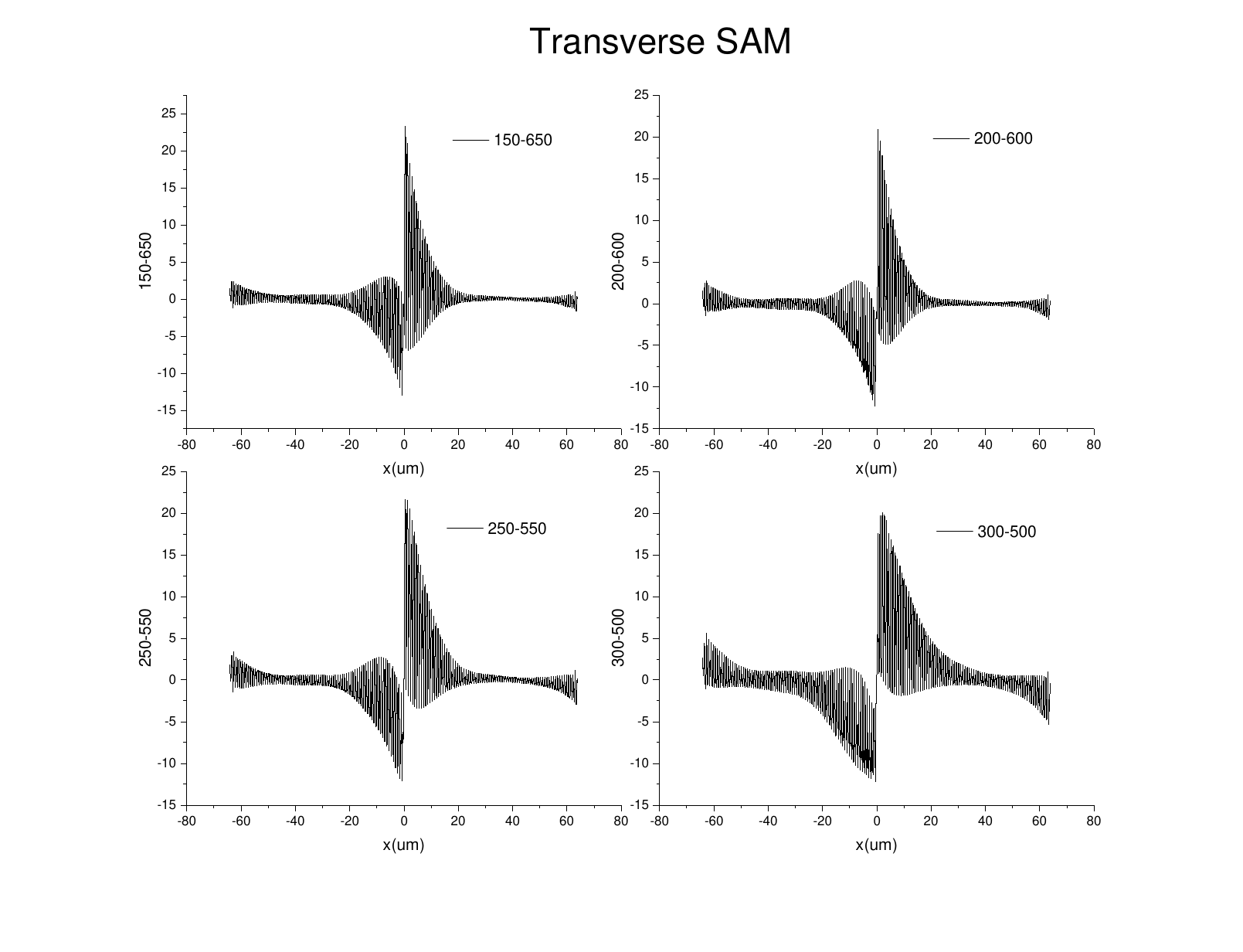}}}
    \qquad
    \subfloat[\centering $\omega_c'$ vs $P_x$ vs SAM]{{\includegraphics[width=0.25\textwidth]{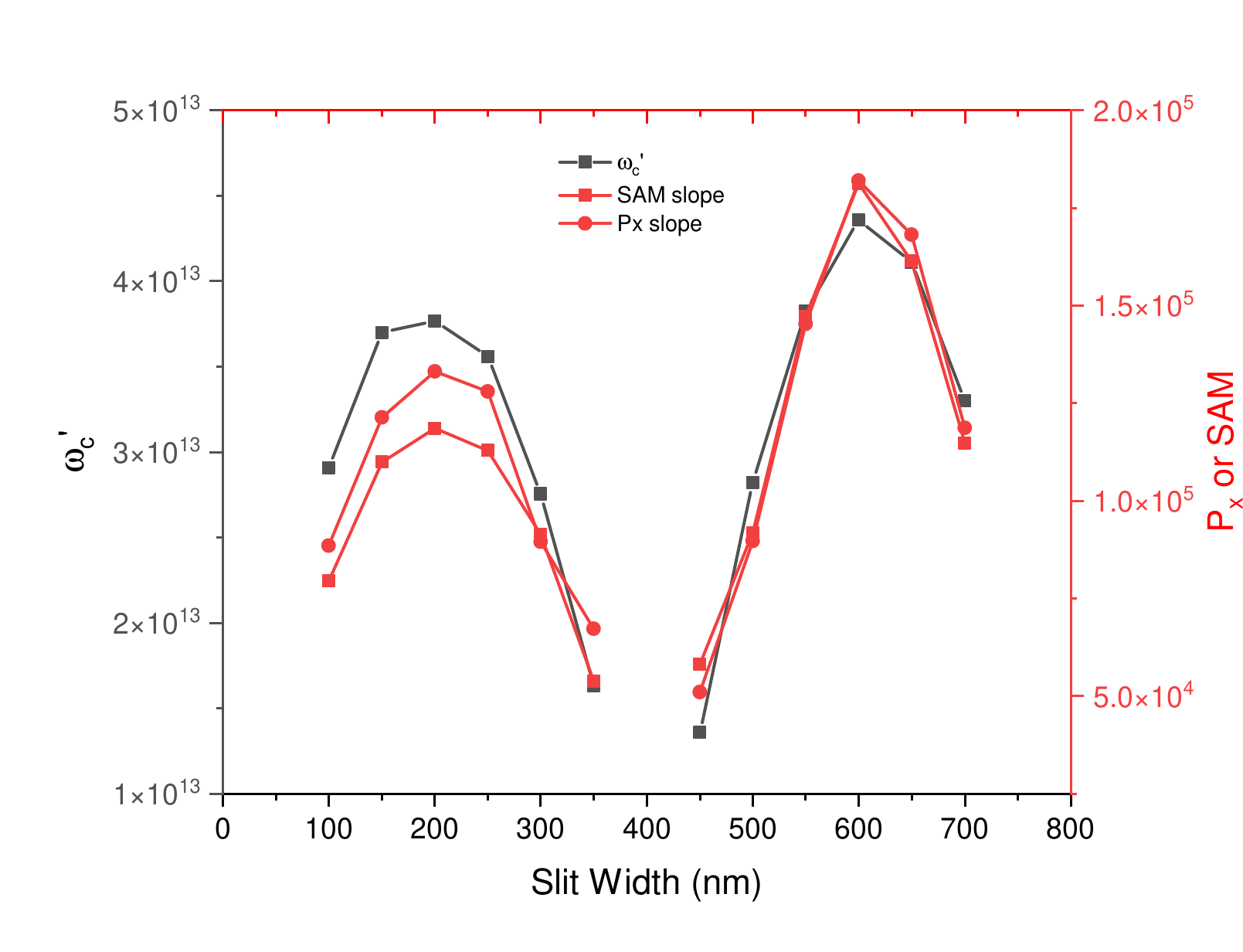}}}
\caption{Poynting vector and transverse SAM simulated for the 200-600nm interface state. (a): Poynting vector. (b): Transverse SAM. The y-axes are in arbitrary unit.  The envelopes follow the form predicted by theory, and the two sides carry opposite signs, demonstrating spin-momentum locking in the topological plasmonic JR state. (c): Comparison between the slopes of $P_x$, SAM, and the field pattern ($\omega_c'$). In section \ref{section: Theory} we predicted their spatial decay should be dominated by the $e^{-m'x}$ term. The similarity of the trends confirms the theoretical prediction.}
\label{fig: sim_poynting_SAM}
\end{figure}

\noindent The field patterns of the interface states can be used to verify our theoretical prediction in section \ref{Poynting and SAM}. The Poynting vectors can be directly obtained from the simulation, while the transverse SAM can be computed from the simulated E and H fields. Using the coordinate system defined in Fig.\ref{fig: schematic diagram and SEM}a with Eqt.\ref{Px} and \ref{interface state transverse SAM}, we expect $P_x > 0$ and $S_{\perp} < 0$ for $x<0$, while $P_x < 0$ and $S_{\perp} > 0$ for $x>0$. Our theory assumes the uncoupled frequency $\tilde{\omega}_0$ is the same on both sides of the grating, which is not generally true~\cite{Cao:14} even for gratings with near-identical band gaps. Nevertheless, we expect the overall directions of both the Poynting vector and transverse SAM to follow our theoretical prediction because they arise from the intrinsic properties of propagating SPPs. Furthermore, as the exponential term is the only decay term in the Hermitian JR solution and remains the dominant decay term for the non-Hermitian fields, we expect the spatial decay of the Poynting vector and transverse SAM in our simulations to largely follow $e^{-2m'x}$. Fig.\ref{fig: sim_poynting_SAM} shows the simulation results for Poynting vectors and SAM. All simulated interface states exhibit similar behavior: the signs of the envelopes agree with theoretical predictions, and the spatial decay trends for both Poynting vectors and SAM correlate with $\omega_c'$, further verifying the predictions in section \ref{subsection: Poynting and SAM}.\\

\section{Experimental Verification}\label{Section: experiment}

\noindent To experimentally validate our theoretical framework, we fabricated a series of 1D rectangular Au plasmonic crystals (PmCs) using focused ion beam (FIB) milling and characterized them with angle- and polarisation-resolved diffraction spectroscopy~\cite{chan2025determinationirreduciblerepresentationshigh}. Scanning electron microscopy (SEM) images, shown in Fig.\ref{fig: schematic diagram and SEM}b, confirm that the fabricated structures closely match the geometry used in simulations (Table \ref{table: width_table}). The samples were characterized using a custom-built Fourier-space optical microscope; We employed two independent experimental methods to determine the coupling constant $\omega_c'$: (1) measuring the band gap of bulk PmCs and (2) imaging the spatial decay of the localized interface state.

\subsection{Bulk Band Structure Measurement}\label{subsection: bulk measurement}

\noindent The band structure of each bulk PmC was measured using angle- and polarisation-resolved diffraction spectroscopy. With p-polarized illumination, the reflectivity was calculated as the ratio of the intensity reflected from the PmC to that from a flat Au reference. Specular reflectivity maps were recorded along the $\Gamma$-X direction by varying the incident angle from $-3^{\circ}$ to $3^{\circ}$ in $0.15^{\circ}$ increments, matching the FDTD simulation parameters. Representative measured band structures are shown in Fig.\ref{fig: band_structures}(e-h).\\

\noindent The measured dispersion relations at larger angles are nearly identical for all samples, as expected for gratings with the same period. Near normal incidence ($\theta \approx 0$), however, the band gap size depends strongly on slit width $W$. As shown in Fig.\ref{fig: band_structures}(e-h), a clear band-inversion signature is observed: the band gap closes and reopens as the slit width crosses the critical value $W \approx 400$nm. These results agree well with the FDTD simulations (Fig.\ref{fig: band_structures}(a-d)). Following the procedure used in the simulations, we extracted the band gap for each bulk grating to determine experimental values of $\omega_c'$, which are plotted in Fig.\ref{fig: omega_c_and_dirac_mass}(b).

\subsection{Field Pattern Imaging}\label{subsection: field imaging}

\begin{figure}[h!]
\centering
    \centering
    \subfloat[\centering Simulated Field Pattern]{{\includegraphics[width=0.22\textwidth]{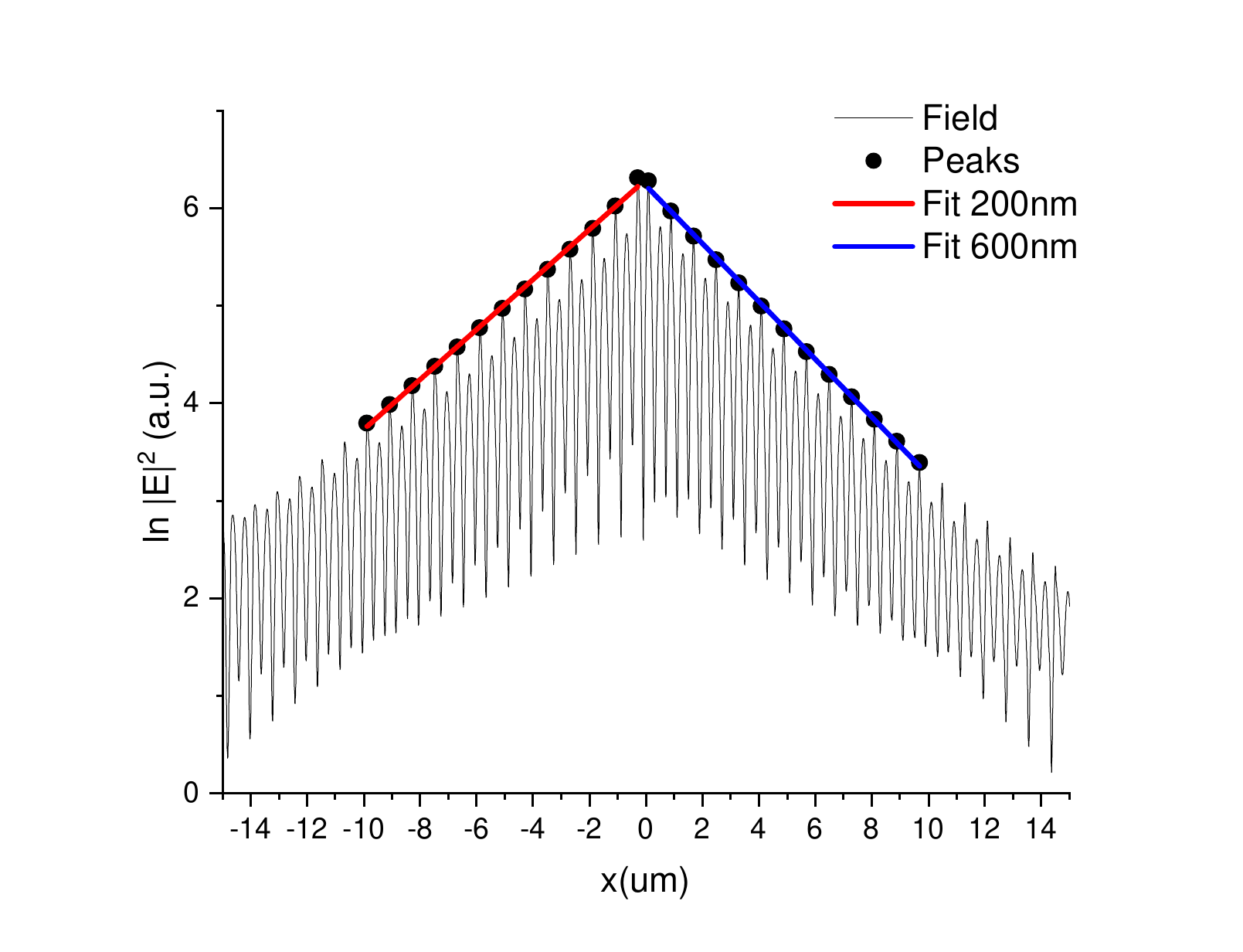}}}
    \qquad
    \subfloat[\centering Real Space Image]{{\includegraphics[width=0.22\textwidth]{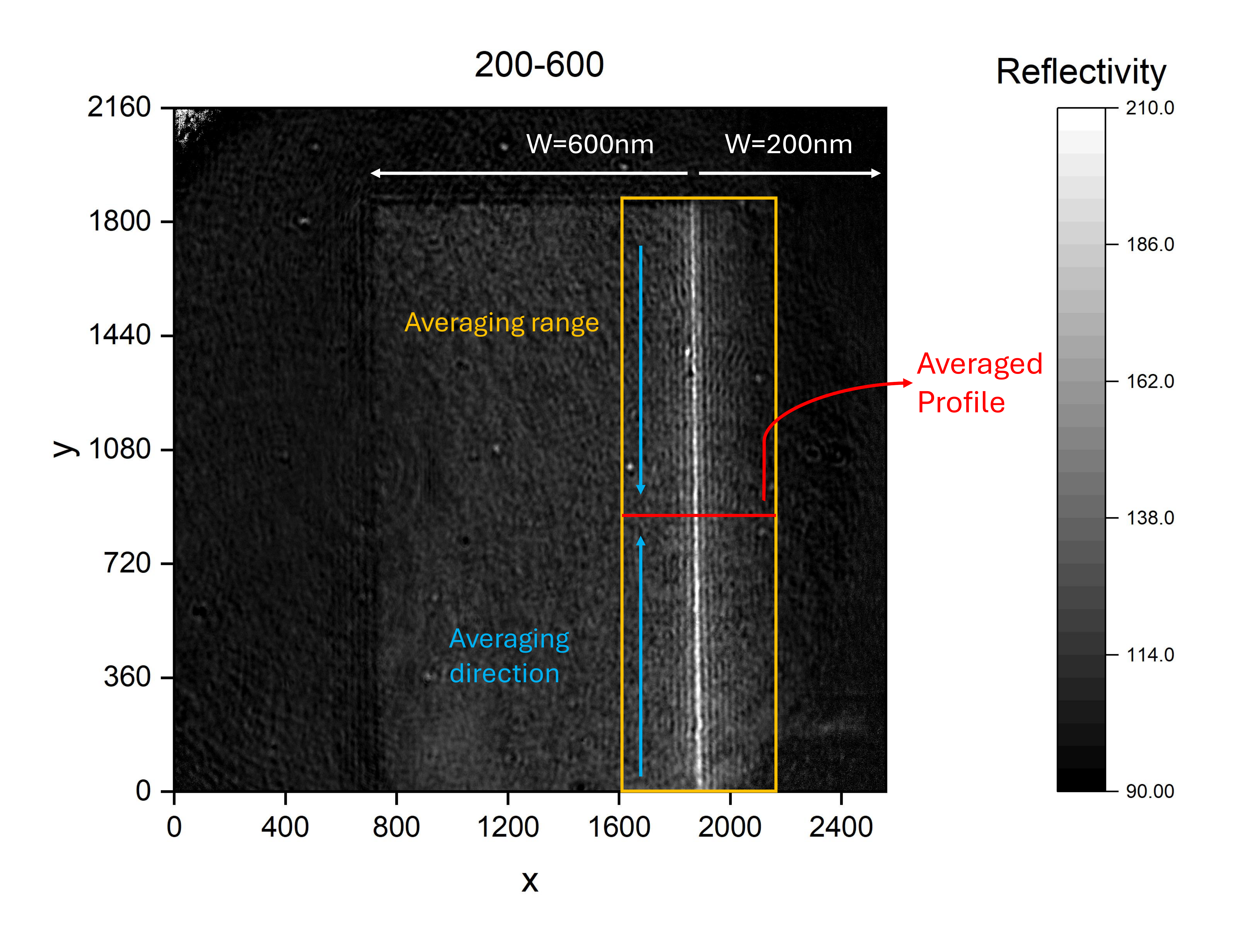}}}
    \subfloat[\centering Averaged Field Intensity Profile]{{\includegraphics[width=0.22\textwidth]{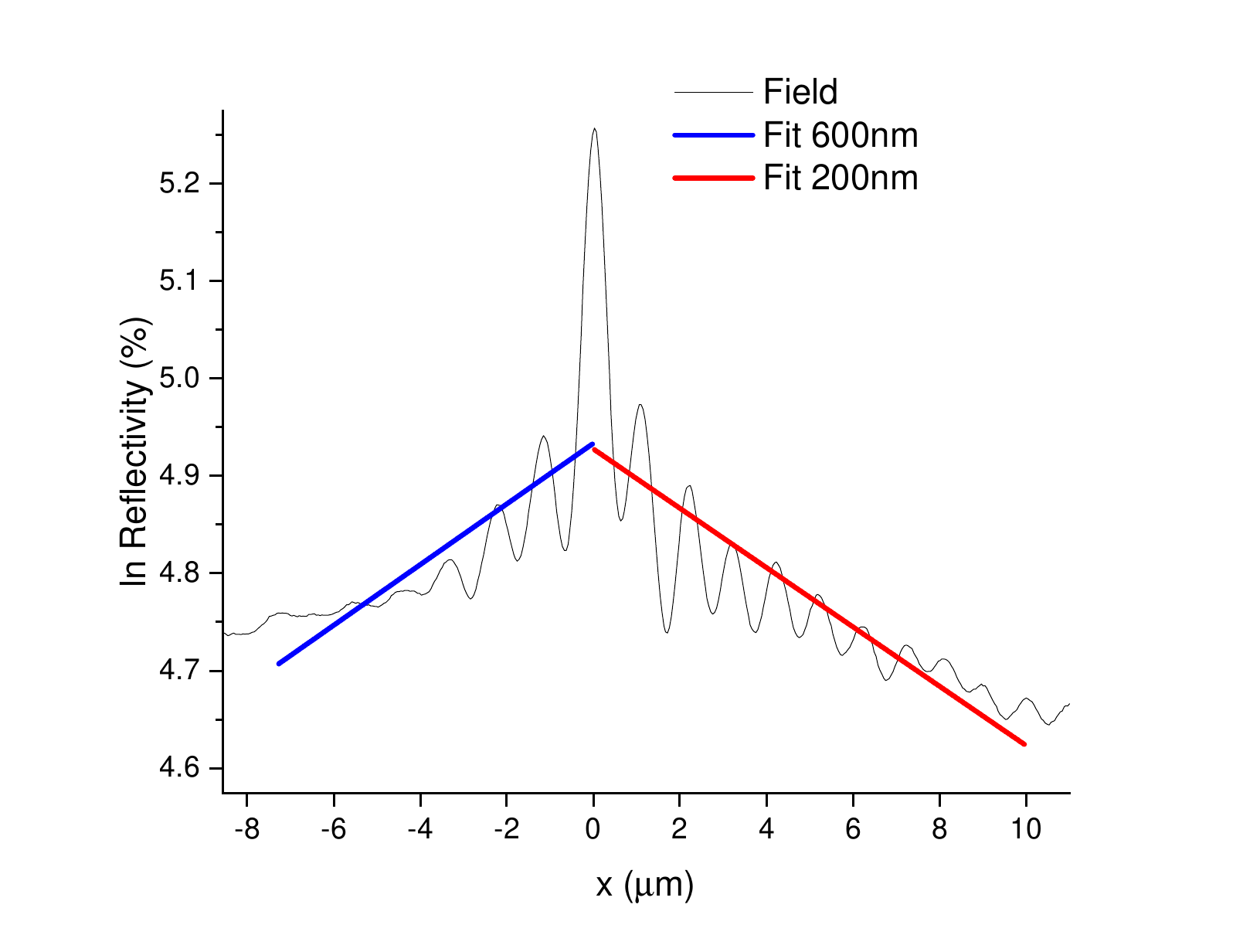}}}
\caption{Simulated near-field patterns plotted on a logarithmic scale. P-polarised light at frequency $\omega_0$ is normally incident on the interface. A spatial monitor records the near-field pattern. As shown in Eqt.\ref{log_E^2}, the envelope of the log-scale intensity should be a straight line. The slope of the linear fit is used to calculate $\omega_c'$. (b): Real-space image obtained using the orthogonal imaging scheme. The interface state emits light at the resonance frequency after excitation with p-polarised light. The signal is averaged along the direction of the slits and used to determine $\omega_c'$. (c): Intensity profiles captured by field-pattern imaging, plotted on a logarithmic scale. The decaying profile was fitted using linear regression; the slope was used to calculate $\omega_c'$ as shown in Fig.\ref{fig: omega_c_and_dirac_mass}(b).}
\label{fig: interface state fields}
\end{figure}

\noindent To image the interface state's field distribution, the microscope was reconfigured for real-space imaging. We employed an orthogonal polarisation detection scheme to isolate the interface-state emission. The sample was illuminated with light polarized at $+45^{\circ}$. The specularly reflected light largely preserves this polarisation, whereas light scattered from the p-polarized SPP interface state is p-polarized. Placing a linear polarizer set to $-45^{\circ}$ in the detection path suppresses the specular reflection, allowing a camera to capture the real-space image of the light emitted by the interface state. The real-space image is shown in Fig.\ref{fig: interface state fields}(b). \\

\begin{figure}[h!]
\centering
    \centering
    \subfloat[\centering Simulated Coupling Constants]{{\includegraphics[width=0.22\textwidth]{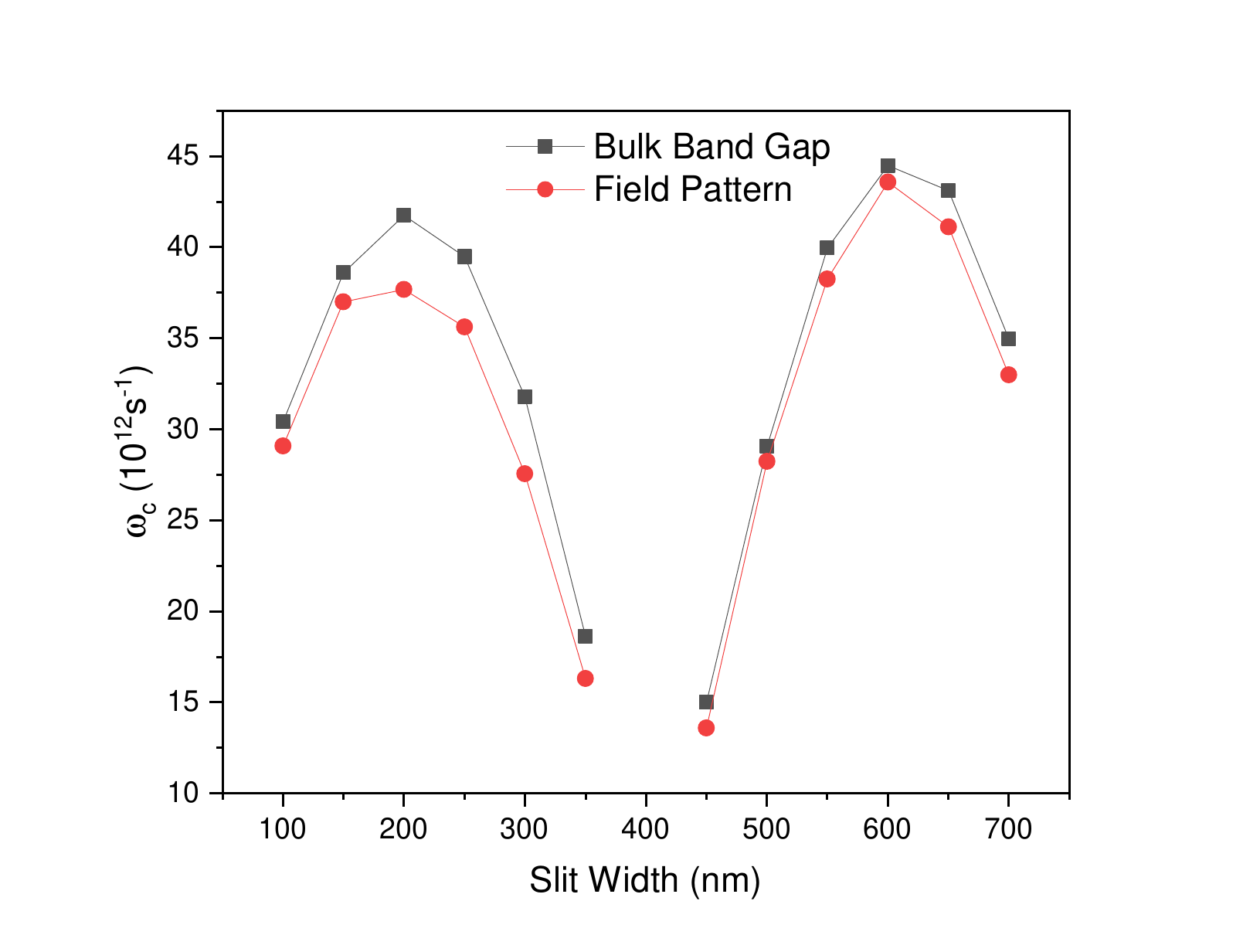}}}
    \subfloat[\centering Measured Coupling Constants ]{{\includegraphics[width=0.22\textwidth]{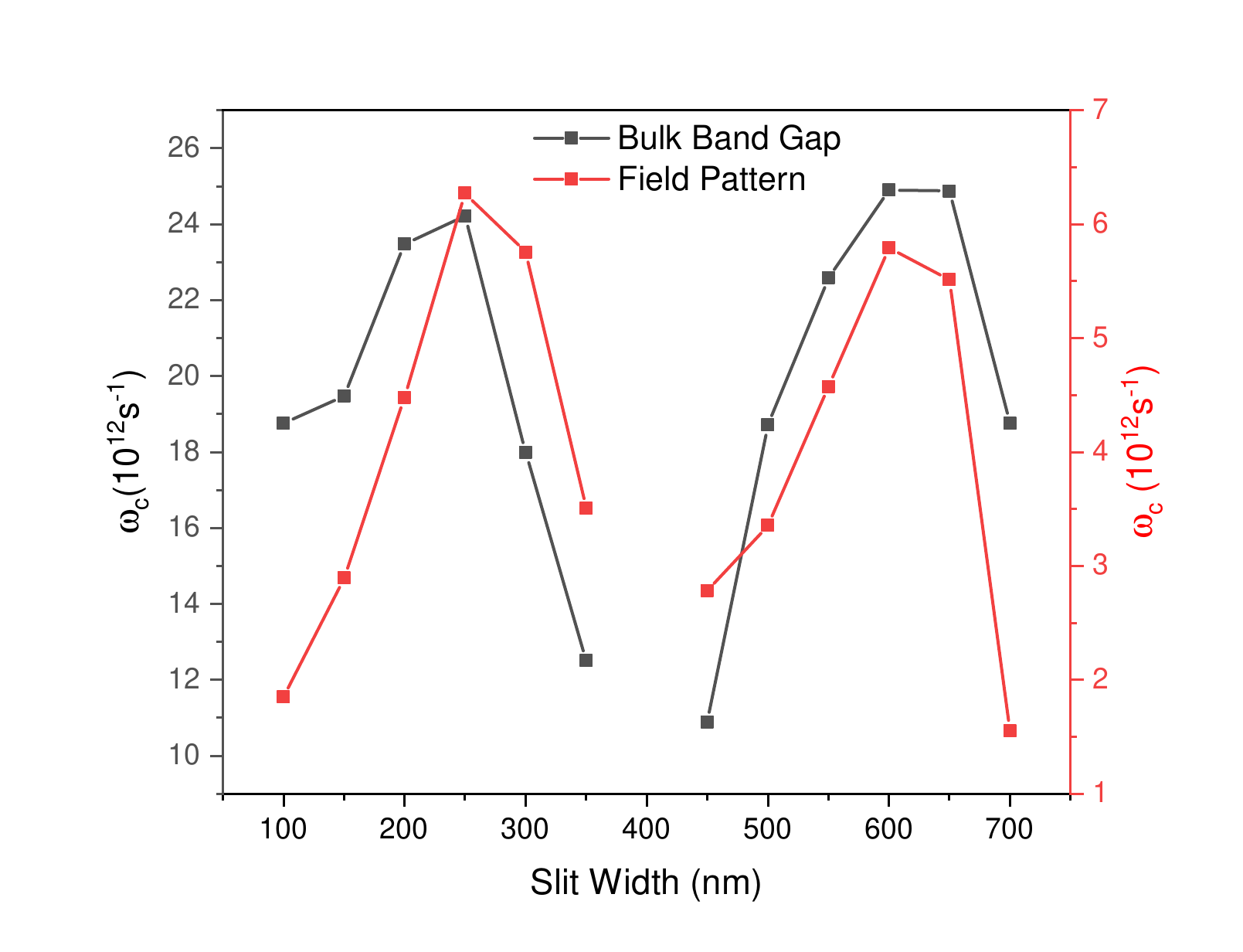}}}
\caption{(a): $\omega_c'$ values derived from field-pattern decay (interface simulation) and from band gaps (bulk simulation). The values from the two methods show nearly identical trends and magnitudes, confirming that the coupling constant of an interface state can be accurately determined from either its field decay or the band gaps of its constituent parts. (b): $\omega_c'$ obtained from bulk band-gap measurements and from field-pattern imaging, showing similar trends to the simulation results.}
\label{fig: omega_c_and_dirac_mass}
\end{figure}

\noindent Intensity profiles perpendicular to the interface were extracted from the captured images and plotted on a logarithmic scale in Fig.\ref{fig: interface state fields}(c). Note that this technique records the far-field radiation pattern, not the near-field intensity as in the simulations. The measured profiles exhibit a near exponential decay, similar to simulations. To robustly extract the decay constant and reduce the impact of experimental noise, we performed a linear fit to the entire logarithmic intensity profile rather than fitting only the peaks of the oscillations. The values of $\omega_c'$ derived from these fits are presented in Fig. \ref{fig: omega_c_and_dirac_mass}(b), showing trends consistent with those obtained from the band-gap measurements.

\subsection{Mode Volume}

\begin{figure}[h!]
\centering
    \centering
    \subfloat[\centering Simulated Mode Volumes]{{\includegraphics[width=0.22\textwidth]{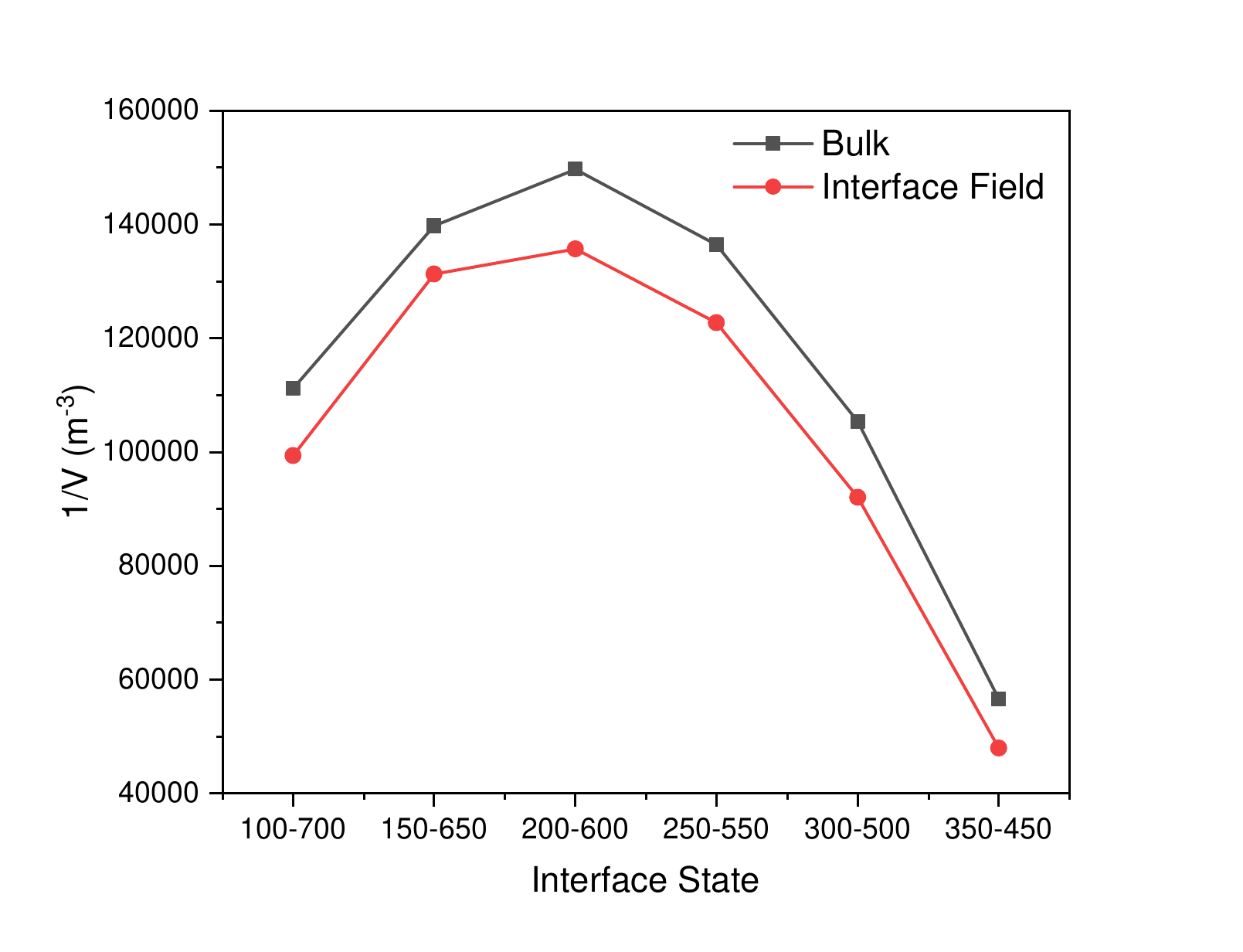}}}
    \subfloat[\centering Measured Mode Volumes]{{\includegraphics[width=0.22\textwidth]{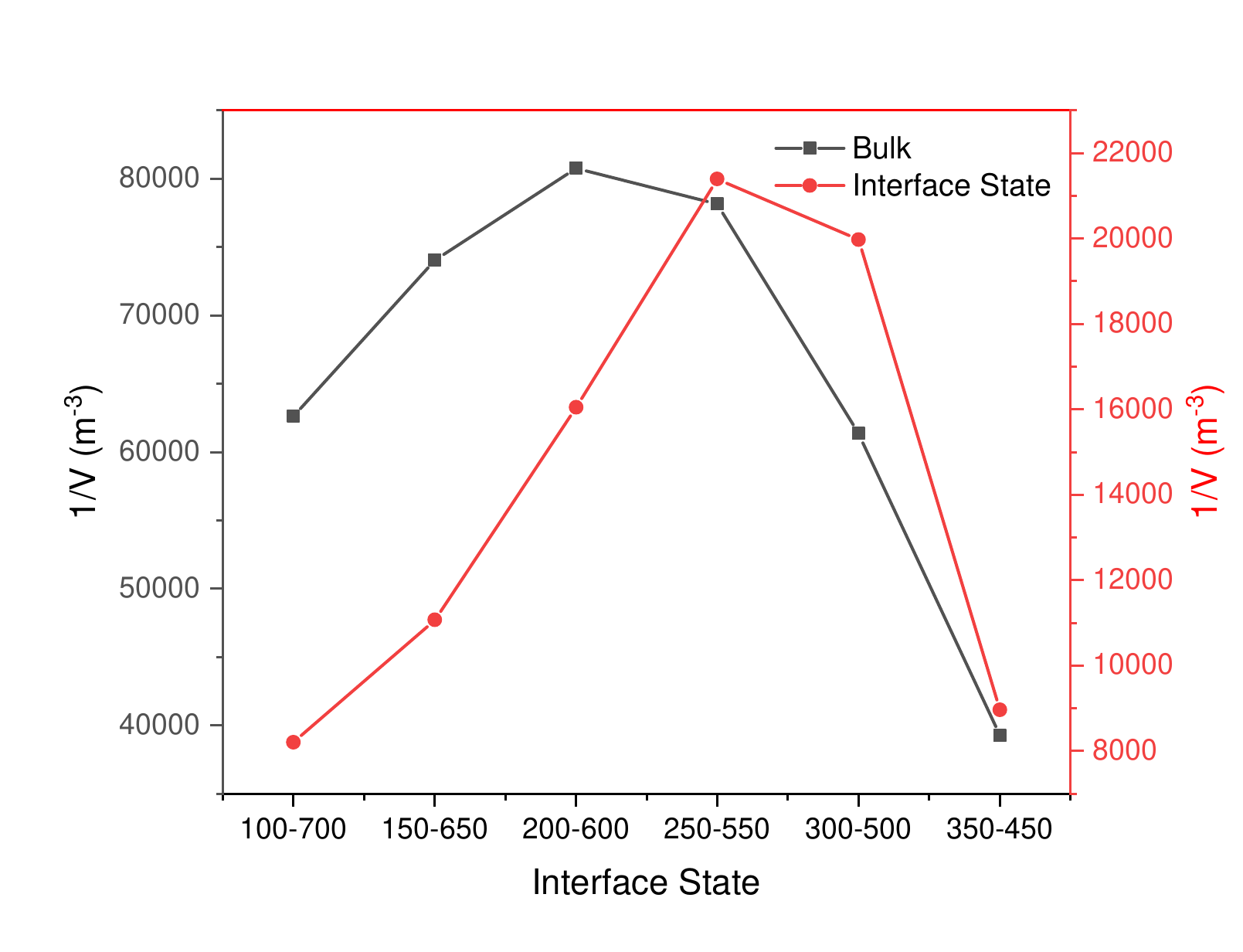}}}
\caption{(a): Inverse mode volume $1/V_m$ calculated using $\omega_c'$ values from both bulk and field-pattern simulations. The results from both methods are nearly identical, showing that larger $\omega_c'$ yields larger $1/V_m$ and thus higher energy confinement. (b) Inverse mode volume $1/V_m$ calculated using the $\omega_c'$ values from both experimental methods. Although the absolute values differ between the methods, the trends are similar and consistent with simulation, confirming the theoretical predictions.}
\label{fig: mode-volume}
\end{figure}

\noindent Having obtained two independent experimental sets of $\omega_c'$, we calculated the corresponding inverse mode volumes $1/V_m$ using Eqt.\ref{dual sided V}. The results are plotted in Fig. \ref{fig: mode-volume}(b). While there is some quantitative disagreement between the two measurement methods—attributable to experimental imperfections—the overall trends agree with the simulation results. Specifically, interface states formed by gratings with larger coupling constants (larger band gaps) yield larger $1/V_m$, indicating stronger energy confinement. This confirms the main conclusion of our theoretical model: the mode volume of a topological interface state is fundamentally governed by the band gaps of its constituent parts.

\newpage

\section{Conclusion}\label{Section: Conclusion}

\noindent In summary, we have established a comprehensive theoretical framework, rooted in spatiotemporal Coupled-Mode Theory (CMT), to elucidate non-Hermitian resonance states in topological guided-mode resonance gratings. This framework remains compatible with the canonical JR state of the Dirac equation while extending into the non-Hermitian regime. A key prediction of our theory is the emergence of spin-momentum locking in the JR solution, arising from the SAM inherent to the evanescent electromagnetic wave, thereby advancing the fundamental understanding of these topological phenomena in photonics. We also derived a closed-form analytical expression for the effective mode volume, showing that it can be minimized by maximizing the photonic band gaps of the constituent gratings.\\

\noindent The theoretical framework was validated through numerical simulations and direct experimental measurements. Crucially, the coupling constant $\omega_c'$ was determined via two independent methods—analysis of the band gaps and of the resonant field patterns—with excellent agreement across theory, simulation, and experiment. This confluence confirms that spatiotemporal CMT provides a robust and predictive model for photonic topological interface states, opening avenues for their systematic design and practical application.\\


\begin{acknowledgements}

\noindent This research was supported by the Chinese University of Hong Kong through Innovative Technology Fund Guangdong-Hong Kong Technology Cooperation Funding Scheme (GHP/077/20GD), Partnership Research Program (PRP/048/22FX), and Seed Program (ITS/245/23).

\end{acknowledgements}


\bibliographystyle{unsrturl}
\bibliography{citations}
\label{LastBibItem}

\end{document}